\documentclass[12pt,preprint]{aastex}
\usepackage{amsmath}
\usepackage{amssymb}
\renewcommand{\mag}{\mbox{$\;$mag}}
\begin{document}
% ******************************************************************
\title{\MakeUppercase{%
     The Tolman Surface Brightness Test for the Reality of the
     Expansion. V. Provenance of the Test and a New Representation 
     of the Data for Three Remote \textsl{HST} Galaxy Clusters}} 

\author{Allan Sandage}
\affil{The Observatories of the Carnegie Institution of Washington,\\
       813 Santa Barbara Street, Pasadena CA, 91101}

%
%%% citation within affil
\affil{Details sire the complications: {\rm old Chinese saying}}

%
% ******************************************************************
%    Abstract
% ******************************************************************
\begin{abstract}
A new reduction is made of the \textsl{HST} photometric data for E
galaxies in three remote clusters at redshifts near $z=0.85$ in search
for the Tolman surface brightness (SB) signal for the reality of the
expansion. Because of the strong variation of SB of such galaxies with
intrinsic size, and because the Tolman test is about surface
brightness, we must account for the variation. In an earlier version
of the test, Lubin \& Sandage calibrated the variation out. In
contrast, the test is made here using fixed radius bins for both the
local and remote samples. Homologous positions in the galaxy image at
which to compare the surface brightness values are defined by radii at
five Petrosian $\eta$ values ranging from 1.0 to 2.0. S{\'e}rsic 
luminosity profiles are used to generate two diagnostic diagrams that 
define the mean SB distribution across the galaxy image. A S{\'e}rsic
exponent, defined by the $r^{n}$ family of S{\'e}rsic profiles, 
of $n=0.46$ fits both the local and remote samples. Diagrams of the 
dimming of the $\langle$SB$\rangle$ with redshift over the range of
Petrosian $\eta$ radii shows a highly significance Tolman signal but 
degraded by luminosity evolution in the look-back time. The expansion
is real and a luminosity evolution exists at the mean redshift of the 
\textsl{HST} clusters of $0.8\mag$ in $R_{\rm cape}$ and $0.4\mag$ in 
the $I_{\rm cape}$ photometric rest-frame bands, consistent with the
evolution models of Bruzual and Charlot. 
\end{abstract}

% ******************************************************************
%     1. INTRODUCTION
% ******************************************************************
\section{\MakeUppercase{Introduction}}
\label{sec:01}
In one of the outstanding ironies in the history of observational
cosmology, Hubble, even in his last years, expressed doubts about the
reality of the expansion. His reasons were based on what he considered
to be anomalies in the correlation of apparent magnitudes and
redshifts (the Hubble diagram) and in his galaxy counts 
\citep{Hubble:34,Hubble:36}, when both are corrected for the effects
of redshifts on the apparent magnitudes. These corrections, the $K$
terms, are calculated by shifting an assumed spectral energy curve of
the mean mixture of galaxy types through the fixed photometric band
pass of the detector. Applying his calculated corrections to his
redshift diagram and his $N(m)$ count data gave Hubble what seemed to
be unacceptable results if the expansion is real.  

     His most direct statements were these.

     (1). ``It is evident that the observed result, [of applying a
blue $K$ correction of $2.94\, z\mag$] is accounted for if the
redshifts are not velocity shifts''.  \dots\ [the data are consistent
but only if] ``the expansion and spatial curvature are either
negligible or zero'' \citep[][p.~542]{Hubble:36}.

     (2). In considering the redshift-distance relation, ``The
inclusion of recession factors [to the magnitudes] would displace all
the points [in the Hubble diagram] to the left [higher redshifts at a
given magnitude], thus destroying the linearity of the law of
redshifts'' \citep{Hubble:37}. 

     (3). \dots\ ``if redshifts are not primarily due to velocity 
shifts \dots\ [then] the velocity-distance relation is linear, the 
distribution of nebulae is uniform, there is no evidence of 
expansion, no trace of curvature, no restriction of the time 
scale.'' [But] ``the unexpected and truly remarkable features are 
introduced by the additional assumption that redshifts [actually 
do] measure recession. The velocity-distance relation deviates 
from linearity by the exact amount of the postulated recession. 
The distribution departs from uniformity by the exact amount of 
the recession. The departures are compensated by curvature which 
is the exact equivalent of the recession. Unless the coincidences 
are evidence of an underlying necessary relation between the 
various factors, they detract materially from the plausibility of 
the interpretation. \dots\ the small scale of the expanding model both 
in space and time is a novelty, and as such will require rather 
decisive evidence for its acceptance'' \citep[][p.~553/4]{Hubble:36}. 

     (4). And finally in his Darwin Lecture in \citeyear{Hubble:53},
only a few months before his death: ``When no recession factors are
included, the law will represent approximately a linear relation
between  redshift and distance. When recession factors are included,
the  distance relation [becomes] non-linear.'' [If no recession factor 
is included] ``the age of the universe is likely to be between 
3000 and 4000 million years, and thus [again with no recession 
factor] comparable with the age of the rock crust of the Earth'' 
\citep{Hubble:53}. (Here concerning the time scale, Hubble clearly 
was factoring in the beginning of the major correction to his 
distance scale that would eventually reach a ratio between 7 and 
10 for the new scale to the old. Clearly, in 1953 he was 
beginning to accept that his value of the Hubble constant must be 
considerably revised downward, based, as it was in 1953, on 
Baade's revision of Hubble's distance to M31 by a scale factor 
close to 2). 

     It is now known that these arguments against a true 
recession are incorrect. Details of why have been variously set 
out elsewhere, the most recent in several reviews 
\citep{Sandage:95,Sandage:98}. The essence of the case is threefold.  

     (A). The $K$ corrections for the effects of redshifts on 
apparent magnitudes used by Hubble are incorrect because he 
assumed a black body spectral energy distribution (SED) of 6000 K 
temperature whereas \citet{Greenstein:38} had shown already in 1938 
that the correct `color temperature' of the bulge of M31 is 
closer to 4200 K. A directly observed template SED for E galaxies 
was not measured until beginning in 1968 
\citep{Oke:Sandage:68,Whitford:71,Schild:Oke:71,Code:Welch:79}, 
and eventually for galaxies of other morphological types.
Examples include the works by \citet{Wells:73}, \citet{Pence:76}, 
\citet*{Coleman:etal:80}, and \citet{Yoshii:Takahara:88}, and 
thereafter by many others in various pass bands, summarized elsewhere 
\citep[][\S~4.2]{Sandage:88}.    

     (B). Hubble's magnitude scales were extrapolations of the Mount
Wilson Catalogue of Photographic Magnitudes 
\citep{Seares:etal:30}, which themselves needed systematic 
corrections fainter than about 16 apparent blue magnitude 
\citep{Stebbins:etal:50,Sandage:01}. Baade had often called Hubble's
approximations ``enthusiastic magnitudes'', not out of derision but as
a tribute to Hubble's skill in devising by practical methods what he
needed in his reconnaissance studies of difficult problems.  

     (C). Hubble's definition of redshift distance as $D = cz/H_{0}$
is the intuitive choice but is incorrect in the standard model of
cosmological parameters. The exact formulation was not made until the
fundamental paper by \citet{Mattig:58}. The difference between
Hubble's assumption and the exact formulation of metric coordinate
distance for different geometries and redshifts is large, as seen in
Figure~4 of \citet{Sandage:98}.

     The consequence of these developments has been that Hubble's
reticence to accept the redshift-distance relation as due to real
recession is no longer valid. There is almost universal acceptance
that the redshift phenomenon which increases linearly with distance is
due to a real recession. Of the several reasons, the most direct is
the excellent agreement in the three time scales of (a) the expansion
age once the Hubble constant is known, (b) the age of the oldest
globular clusters in the Galaxy, and (c) the age of the chemical
elements. All three average about $14 \times 10^{9}$ years.  

     Other less direct tests, less direct because they require more
complicated assumptions concerning a hot early universe giving arise
to the $3^{\circ}$ MWB, include the observation that the temperature 
of the MWB is hotter at high redshift as we look back in time
\citep{Songaila:etal:94,Srianand:etal:00,Molaro:etal:02},
the acoustic waves in the correlation of the fluctuation in the MWB,
and, of course the MWB itself. Although in a minority, there are still
astronomers who offer alternate explanations of these latter tests in
their questioning of the standard model of a hot early universe; yet
the time scale agreement remains the decisive test \citep{Sandage:68}
on which there is now such a large literature as to deny an adequate
summary here.        

     Nevertheless, a true expansion where the redshifts are
cosmological, not due to ``some unknown law of nature'' as favored by
Hubble, is itself such a remarkable proposition that decisive proofs
are still of interest, even if only as academic curiosities now.  

     The Tolman surface brightness test is particularly interesting
because its principle is so clear as to give a major predicted
difference in observational data between an expanding manifold and a
stationary one where the redshift would then be due to ``an unknown
law of nature''. \citet{Tolman:30,Tolman:34} discovered  
the effect that the surface brightness of a ``standard'' radiating 
object that is receding with redshift $z$ will be fainter than a 
similar stationary ``standard'' object at rest by $(1+z)^{4}$. 
However, if the manifold is stationary but nevertheless has a redshift
due to an ``unknown law of nature'', the factor is only $(1+z)$. 
The test, as set out in theory by Tolman, is described 
again in \citet{Hubble:Tolman:35} and is derived heuristically 
elsewhere \citep[eg.][]{Sandage:61,Sandage:74,Sandage:95}. 

     The test, simple in principle is difficult in practice because
there are no simple ``standard'' galaxies to compare with each other,
one at high redshift and the other at small. The difficulty of using E
galaxies for the test has been discussed elsewhere 
(\citealt*{SP:90a,SP:90b,SP:91}; \citealt*{SL:01}; 
\citealt*{LS:01a,LS:01b,LS:01c}, hereafter LS01a,b,or c) and is not
repeated here.

     The first attempt at the test was made by \citeauthor*{SP:90a}
using ground based data in the references just cited
(hereafter \citeauthor{SP:90a},b; \citeauthor{SP:91}).
The purpose of the present paper is to continue the discussion made in
a second attempt of the test by \citeauthor*{LS:01c} using \textsl{HST}
data (\citeauthor{LS:01c}) by showing more directly that a strong
Tolman signal exists in three remote clusters observed with \textsl{HST} 
by Oke, Postman, \& Lubin.
Their photometric data are published in papers by various permutations
of the order of the authors \citep{OPL:98,PLO:98,PLO:01,LPO:98,LPO:01}.
The analysis by \citeauthor{LS:01c} remains valid that the Tolman
signal exists, but the reductions and the representation of the data
are different here.

% ******************************************************************
%     2. THE PROVENANCE OF THE TOLMAN TEST USING HST
% ******************************************************************
\section{\MakeUppercase{%
         The Provenance of the Tolman Test Using \textsl{HST}}}
\label{sec:02}
Extensive planning related to constructing and deploying a large space
telescope had its beginnings in the early 1960s, although the concept
had already been set out 20 years earlier \citep{Spitzer:46} in the
United States and two decades earlier by \citet{Oberth:23} in
Germany. The project, accepted by NASA in the late 1950s, was first
called the \textsl{LST} for Large Space Telescope, but more
appropriately known to the underground as the Lyman Spitzer Telescope
because Spitzer had done so much to promote the idea.  

     Many planning sessions occurred to define the scientific goals,
and the means to achieve them. By the mid 1960s a NASA department for
the \textsl{LST} project was formed with Nancy Roman as its head. The
chief scientist had also been appointed, the first of which was Charles
O'Dell. In these early days of planning and of selling the project to 
the astronomical community and especially to Congress for funding, the
importance of Roman and O'Dell cannot be overstated.     

     When it became apparent that an \textsl{LST} could be built and
operated remotely, based on the experience and successes of the
Orbiting Astronomical Observatories project led by the Grumman
Aerospace Corporation with their chief engineer F.~P.~Simmons as
coordinator, serious scientific and technical workshops were convened
to advance the project. 
One such conference was held in 1974 at the Park-Sheraton Hotel
in Washington DC, organized by Simmons, then at the
McDonald-Douglas Astronautics Company. In a three day meeting,
astronomers and engineers made presentations to Roman, O'Dell, and the
astronomical and technical community on a variety of scientific
projects ripe for the \textsl{LST}. Space engineers from 12 aerospace
companies also discussed the means to achieve the goals. The
astronomers giving papers were Spitzer, J.~L.~Greenstein, I.~King,
E.~M.~Burbidge, L.~W. Fredrick, G.~Neugebauer, G.~Herbig, J.~Bahcall,
Harlan Smith, and the writer. 

     Together with other projects for \textsl{LST} in observational
cosmology, I discussed the requirements of spatial resolution and
input flux needed to make the Tolman surface brightness test. Details
were set out for a particular form of the test, comparing the ratio of
a suitably defined isophotal diameter (the angular diameter to a
particular isophotal level) in an E galaxy to a suitably 
defined metric diameter (the angular diameter to a given number of
parsecs) for the same galaxy. I concluded that the 
\textsl{LST} would be an ideal instrument to make the test with its
proposed 3-meter aperture and a spatial resolution of 0.05 to 0.1 arc~sec. 
Of course, not known at the time was the way that 
surface brightness over the image of E galaxies at some standard
diameter varies systematically with intrinsic size and absolute
magnitude, or indeed how to define a series of ``standard diameters''
within which to measure the surface brightness. These complications
were to be encountered later. Nevertheless, the project with
\textsl{LST} was set out as a serious observing plan that was ideal
for the capabilities of the proposed telescope. It has worked out that
way, where the Tolman test with \textsl{HST} has been done
(\citeauthor{LS:01a},b,c), 25 years beyond the initial
proposal for a space experiment.    

     The photometric data for three remote clusters observed by 
Oke, Lubin, and Postman, cited earlier as 
\citealt{OPL:98,PLO:98,PLO:01,LPO:98,LPO:01},
and reduced in the way described in \citeauthor{LS:01a},b,c
are again used here. What is different in the analysis is threefold.

     (1). A more comprehensive calibration of the local data on E
galaxy surface brightness as function of intrinsic radius is made in
\S~\ref{sec:03}, reaching to smaller radii than was done in 
\citeauthor{LS:01c}. This is necessary because the galaxies in the
\textsl{HST} clusters are at the faint end of the luminosity function
of the first few ranked galaxies in the calibrating sample taken from
\citet*[][hereafter PL95]{PL:95} and from Table~1 of
\citeauthor{SP:91} used in a  first study of the test.
This new calibration of the $\langle$SB$\rangle$-radius correlation
avoids a less certain extrapolation to the necessary smaller radii for
the three remote clusters than we made in \citeauthor{SL:01} 
(Table~3 there).  

     (2). The variation of surface brightness across the E galaxy
image for particular \citet{Petrosian:76} sizes, for both the local
calibrating galaxies and the remote sample, is explicitly shown as a
function of intrinsic Petrosian radii. The data are divided into
groups of different mean intrinsic radii so as to isolate the large
systematic variation of the mean surface brightness with radius (or
absolute magnitude for $M$ brighter than $M_{V} = -20$,  
eg. see Figs.~1 and 2 of \citeauthor{SP:91}). This dependence of the 
$\langle$SB$\rangle$ zero point on intrinsic radius at a given
standard size degrades a raw Tolman signal and must be calibrated
out. Much of \S~\ref{sec:03} is devoted to determining these
properties of the S{\'e}rsic profiles.  

     (3). We show in \S~\ref{sec:03:d} that the S{\'e}rsic profile
with an exponent of 0.46, rather than a de~Vaucouleurs value of 0.25, 
fits our local sample over its range of intrinsic radii. The 
systematic deviation of the profile exponent from a 
de~Vaucouleurs $r^{0.25}$ value with intrinsic radius is strong and is
mapped. The variation of $n$  with radius is the same as discovered 
by \citet[][Fig.~2]{Binggeli:Jerjen:98}.

% ******************************************************************
%     3. PROPERTIES OF SERSIC LUMINOSITY PROFILE FUNCTIONS IMPORTANT 
%                         FOR THE TOLMAN TEST
% ******************************************************************
\section{\MakeUppercase{%
         Properties Of S{\'e}rsic Luminosity Profile Functions Important 
         For The Tolman Test}}
\label{sec:03}
%
% ******************************************************************
%  3.1. The Petrosian Radius Function
% ******************************************************************
\subsection{The Petrosian Radius Function}
\label{sec:03:a}
In the original 1974 proposal, the test was to be made by comparing
the ratio of isophotal to metric diameters for E galaxies at zero
redshift to those with redshift $z$. For the high redshift galaxies, the 
ratio of $r$(isophotal) to $r$(metric) at a given metric diameter 
should decrease with increasing redshift if a Tolman signal
exists. However, the test is difficult to apply because the metric
diameters depend on the world model. We would also have to apply the 
$K$ correction for the effect of redshift on the isophotal surface
brightness values. We wish to avoid these problems at this stage in
the data reductions. A formulation of the test using Petrosian radii
which avoids these problems is used here instead.

     Some of the remarkable properties of the Petrosian radius
function have been discussed elsewhere (\citeauthor{SP:90a}; 
\citealt{Kron:95}; \citealt{Sandage:95}; \citeauthor{SL:01}) and are
only summarized here.

     Two equivalent definitions are these. (1) A Petrosian radius is a
distance from the center of an image where the surface brightness
(SB), averaged over the image that is interior to that radius is
greater than the profile SB at that radius by a fixed number of
magnitudes. (2) It can be shown 
(\citealt{Djorgovski:Spinrad:81}; \citeauthor{SP:90a}) 
that this is the same number that is calculated from the slope of the
growth curve, as,
\begin{equation}
         \eta\ (\mbox{in mag}) = 2.5\log (5 d\log r/d\mag),       % (1)
\label{eq:01}
\end{equation}
where the growth curve is the summed intensity (in magnitudes) out to
that $r$ and its slope at $r$ is $d\mag/d\log r$. Expressed as the
slope in intensity units, as in \citeauthor{PL:95},
$\alpha (= d\log L/d\log r)$, which is used by some authors.
Of course, $\eta (\mbox{in mag})=2.5\log 2/\alpha$.

     The Petrosian $\eta$ function for radii was used throughout the 
first and second attempts to make the Tolman test 
(\citeauthor{SP:90a},b, \citeauthor{SP:91}; \citeauthor{SL:01}
and \citeauthor{LS:01a},b,c).

% ******************************************************************
%  3.2. The Variation of Petrosian eta Radii Along the Hubble
%       Sequence From E to Sm Types
% ******************************************************************
\subsection{The Variation of Petrosian \boldmath{$\eta$} Radii Along
            the Hubble Sequence From E to Sm Types}
\label{sec:03:b}
\begin{figure}[t]
   \epsscale{0.85} % for Figs. 1-5, 8-15
   \plotone{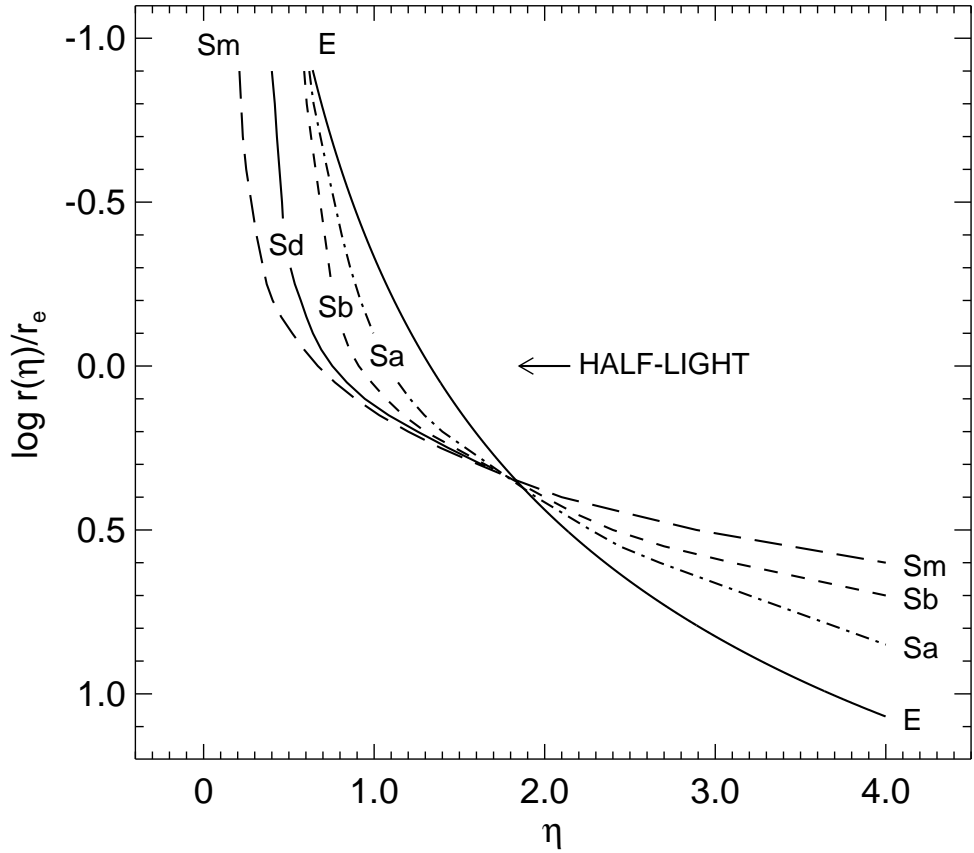}
  \caption{The variation of Petrosian $\eta$ radii with the log of the 
            ratio of the radius to the half-light radius for different 
            galaxy types. The data are from Table~\ref{tab:01}.}
\label{fig:01}
\end{figure}

It is of interest to display the $\eta$ function at various ratios of
radii to the effective (the half-light) radius, $r_{e}$. Because the
luminosity profiles of galaxies are so different along the Hubble
sequence, there is a systematic variation of $\eta$ with morphological
type, as calculated using equation~(\ref{eq:01}) using the growth
curves from Table~11 of the Third Reference Catalogue  
\citep*[][the RC3]{RC3}. The results are listed in 
Table~\ref{tab:01} as function of $\log r/r_{e}$, and shown in
Figure~\ref{fig:01}. The T coding for the morphological types is the
same as in the \citeauthor{RC3}. The even numbered columns of
Table~\ref{tab:01} are the growth curve magnitudes. The odd numbered
columns are the $\eta$ radii at the listed $\log r/r_{e}$ values.   

% ******************************************************************
%  Figure 1: Variation of Petrosian eta radii with log r/r_e
% ******************************************************************

% ******************************************************************
%  Table 1: Growth-Curve Magnitudes and Petrosian eta Radii as 
%           Function of log r/r_e For Various Hubble Types and 
%           for the de Vaucouleurs r^{1/4} law. 
% ******************************************************************

     In the next subsection we show that the entries in 
Table~\ref{tab:01} for the E ($T=-5$) galaxies is close to, but not
identical with the de~Vaucouleurs $r^{0.25}$ profile. We also show
there that the $\eta$ values for the Sm (T = 9) galaxy type is nearly
identical with the exponential intensity profile whose S{\'e}rsic
exponential index is $n=1$.

% ******************************************************************
%  3.3. The eta, log r/r_e and log r/r(eta=2) Relations for the Family
%       of Sersic Luminosity Profiles for E Galaxies} 
% ******************************************************************
\subsection{The \boldmath{$\eta, \log r/r_{e}$} and 
            \boldmath{$\log r/r(\eta=2)$} Relations for the Family
            of S{\'e}rsic Luminosity Profiles for E Galaxies} 
\label{sec:03:c}
It is now well known (citations later in this section) that the 
de~Vaucouleurs $r^{0.25}$ luminosity profile only fits E galaxies near
the bright end of the E galaxy luminosity function that themselves are
not cD subtypes \citep{Oemler:74,Oemler:76}. \citet{Sersic:68} 
generalized the de~Vaucouleurs $r^{0.25}$ profile with the function,  
\begin{equation}
      \log [B(r/r_{e})/B(r_{e})] = -b_{n}[(r/a)^{n} -1],  % (2) 
\label{eq:02}
\end{equation}
where $a$ is a fixed radius to make the radius factor scale free, $n$
is the S{\'e}rsic exponent, and $b_{n}$ is calculated to make the
half-light radius occur at the value of $a=r_{e}$. The de~Vaucouleurs
profile is a special case with $n=0.25$ and $b_{n}=3.33$.  

     The adopted $b_{n}$ values, listed in Table~\ref{tab:02}, were
calculated by numerical integration of equation~(\ref{eq:02}) to
generate growth curves in $r/a$, such that $a=r_{e}$ for each of
S{\'e}rsic profiles with $n$ values of 0.15, 0.2, 0.25, 0.4, 0.46, 0.6,
1.0, and 1.5. The growth-curve magnitudes and Petrosian $\eta$ values
in Table~\ref{tab:03} are at the listed $r/r_{e}$ values for these $n$
values. The listings for $T=-$5, and $n=0.25$ and 0.46 are the same as
in Table~\ref{tab:01}.  

% ******************************************************************
%  Table 2: Calculated b_n Values for the Sersic Family of
%           Luminosity Profiles.  
% ******************************************************************
%  Table 3: Growth Curves and eta Values for the Sersic Function at
%           the Listed log r/r_e Ratios.    
% ******************************************************************

\begin{figure}[t]
   \plotone{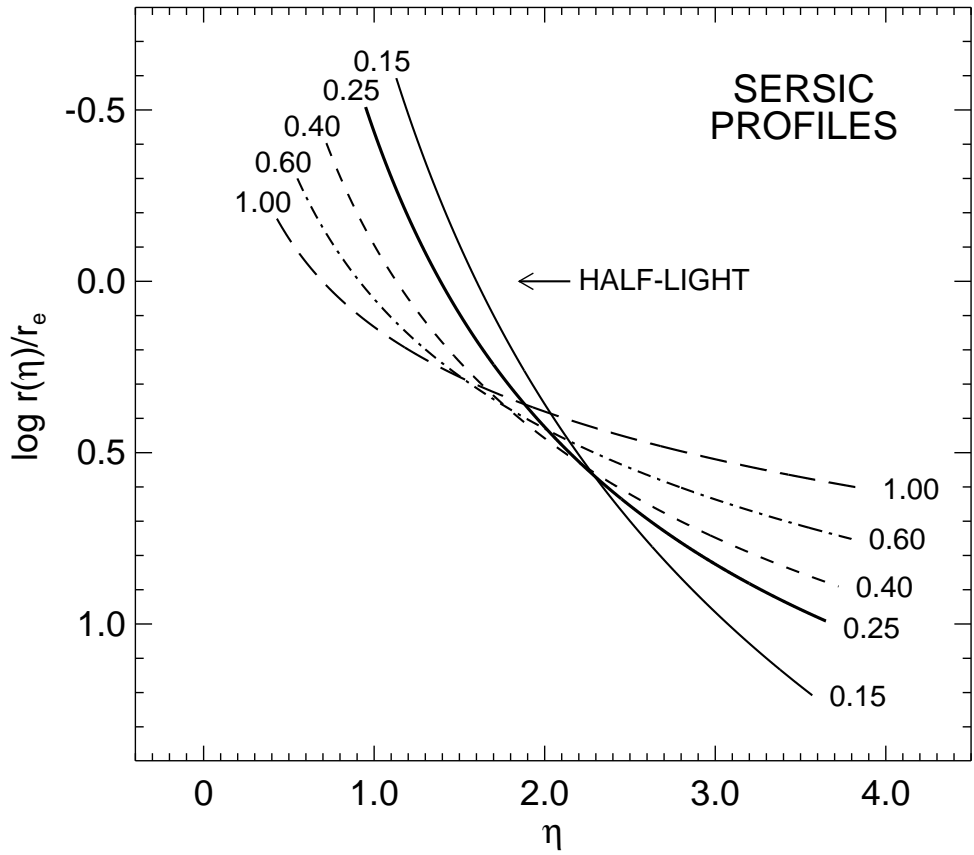}
   \caption{Relation between $\log r/r_{e}$ and $\eta$ for the family of 
            S{\'e}rsic profiles with different S{\'e}rsic exponents. Data are from 
            Table~\ref{tab:03}. Note the similarity to Figure~\ref{fig:01}
            for the different Hubble types.}
\label{fig:02}
\end{figure}

     Figure~\ref{fig:02} shows the variation of $\eta$ with $r/r_{e}$
for the various $n$ values from Table~\ref{tab:03}. If all E galaxies
had a fixed luminosity profile, such as that for the de~Vaucouleurs
$r^{0.25}$ special case, Table~\ref{tab:03} would be germane only to
the E galaxy types for the $T=-5$ column. However, in papers cited
below, it has become known that the S{\'e}rsic $n$ exponent varies
systematically with absolute magnitude in E and dE galaxies, ranging
from 0.15 to 1.5 as the absolute magnitude varies between $M_{B}=-23$
and $-13$.   

% ******************************************************************
%  Figure 2: Relation between log r/r_e and eta
% ******************************************************************

     The progression of $\langle$SB$\rangle$ with absolute magnitude
was discovered by many authors 
\citep[eg.][]{%
Oemler:73,Oemler:74,  
Kormendy:77,Kormendy:87,
Strom:Strom:78a,Strom:Strom:78b,Strom:Strom:78c,
Michard:79,
Thomsen:Frandsen:83,
BST:84,
Choloniewski:85,
Schombert:86,
Ichikawa:etal:86,
Caldwell:Bothun:87,
Djorgovski:Davis:87,
Ferguson:Sandage:88,
Impey:etal:88,
Caon:etal:93}
and we suspect many others. If the deviation from the classical Hubble
or de~Vaucouleurs E galaxy profiles can be described by S{\'e}rsic
functions, which we prove in the next section, it is this deviation
that is listed in Table~\ref{tab:03}.\footnote{% 
   This paper was finished in first draft in February, 2009 and was at
   the stage of first revision when Francois Schweizer drew my
   attention to the comprehensive account by \citet{Kormendy:etal:09}
   of using S{\'e}rsic profiles for E galaxies. These authors develop, as
   we do here, various useful properties of the S{\'e}rsic $r^{n}$
   functions and conclude, as we also, that S{\'e}rsic profiles with their
   range of $n$ values fit the observed E galaxy surface photometry
   remarkably well. However, the treatment differs in the two
   works. We develop the properties of the S{\'e}rsic functions in terms
   of Petrosian $\eta$ radii to define homologous regions of the image
   for galaxies of different absolute magnitude. 
   \citeauthor{Kormendy:etal:09} do not use Petrosian radii because
   they have no need in their study of origins to define similar
   points over an E galaxy image, which we must for the Tolman
   test. Such points are naturally given by Petrosian radii. The
   sometimes parallel conclusions, here and by
   \citeauthor{Kormendy:etal:09}, of the properties of S{\'e}rsic
   functions, agree.}

     But before giving the proof, we first rearrange
Table~\ref{tab:03} to use $\eta$ rather than $r/r_{e}$ as the
independent variable. The table is also renormalized to change the
fiducial radius from the half-light radius to the radius at $\eta=2$,
hereafter called $r(2)$. The change is made because it will be more
convenient in \S~\ref{sec:05} to use $r/r(2)$ to compare surface
brightness levels at various points rather than $r_{e}$. To this end,
Table~\ref{tab:04} is reconstituted by a graphical interpolation of
Table~\ref{tab:03}, and then defines the radius at $\eta=2$ to be the
zero point of the coordinate radius. Figure~\ref{fig:03} shows the
data from Table~\ref{tab:04} for the ratio of the various $r(\eta)$
radii to $r(2)$. This diagnostic diagram will be often used later to
compare observational data for local and the \textsl{HST} cluster
galaxies.   

% ******************************************************************
%  Figure 3: Correlation of eta with log r(eta)/r(2)
% ******************************************************************

% ******************************************************************
%  Table 4: log r/r(eta=2) at Various eta Values for the Sersic 
%           Family from Table 3.
% ******************************************************************

% ******************************************************************
%  3.4. Oemler Profiles
% ******************************************************************
\subsection{Oemler Profiles}
\label{sec:03:d}
But before comparing the entries in Tables~\ref{tab:03} and
\ref{tab:04} with observed E galaxy profiles, it is of interest to
compare the S{\'e}rsic profiles with those of
\citet{Oemler:74,Oemler:76}. Oemler's function embrace a family based
on a modified Hubble profile with various strengths of a luminosity
turndown at large radii. The Oemler profiles have the equation, 
\begin{equation}
   B(r)/B_{0}=\frac{\mbox{exp}(-r/\alpha)^{2}}{(1 + r/\beta)^{2}}  % (3)
\label{eq:03}
\end{equation}               
where $\alpha$ and $\beta$ are fitting parameters. This function is
illustrated in \citeauthor{SP:90a} where the resulting family,
relative to the Hubble profile, is shown in Figures~2 and 3 of that 
paper. The profiles are generated by varying the $\alpha/\beta$ ratio,
and is made scale free by using $r/\beta$ as the radius measure.

     Note that the denominator on the right of equation~(\ref{eq:03}) 
is the Hubble law. The numerator is the added decay factor making the
integrated growth curve finite, which the raw Hubble law does
not. The similarity and the differences with the S{\'e}rsic family that is
used here is in Figure~\ref{fig:04}, taken from Table~\ref{fig:05}
where the $\eta$, $r/r(\eta=2)$ Oemler profiles are listed for
$\alpha/\beta$ ratios of 10, 30, 60, and 100. The $T=-5$ relation is
copied from Table~\ref{tab:01} but is modified in an obvious way to be
a function of $r/r(2)$ rather than $r/r_{e}$. Three Oemler profiles
from Table~\ref{tab:05} are shown and are compared with the $T=-5$
profile from Table~\ref{tab:01} and with the $n=0.46$ S{\'e}rsic profile
from Table~\ref{tab:03}. The fit of the $T=-5$ data with the 
Oemler profile for $\alpha/\beta=30$ for $\eta$ between 1 and 2 is
excellent, after which the fit must be made to an $\alpha/\beta$ ratio
of 100 for $\eta$ between 2 and 3.5. 
The fit of the $n=0.46$ S{\'e}rsic profile is excellent for an Oemler
profile of $\alpha/\beta=10$ for $\eta<2$, after which an
$\alpha/\beta$ ratio of 30 must be used.
Table~\ref{tab:05} has been calculated in an obvious way from
equation~(\ref{eq:01}) using Table~1 of \citeauthor{SP:90a} for the
slope of the Oemler growth curves.   

% ******************************************************************
%  Table 5: log r/r(eta=2) as Function of eta and the alpha-to-beta 
%           Ratio for Four Oemler Profiles and the Observed T = -5 and 
%           a Sersic Profile With n=0.46.
% ******************************************************************

% ******************************************************************
%  Figure 4: Oemler profiles compared with observed T=-5 galaxy profiles
% ******************************************************************

     Our reason for discussing the Oemler profiles here is that they
are a natural generalization of the Hubble single parameter profile
that was standard for many years, and it was the Hubble profile with
its scale factor, $a$, that was discussed in proposing the Tolman test
at the \textsl{LST} meeting in 1974.

     Return now to the S{\'e}rsic profiles to display the surface 
brightness-$\eta$ properties of the S{\'e}rsic function.

% ******************************************************************
%  3.5. Surface Brightness vs. eta for Sersic Profiles
%       Normalized to eta=2
% ******************************************************************
\subsection{Surface Brightness vs. \boldmath{$\eta$} for S{\'e}rsic
            Profiles Normalized to \boldmath{$\eta=2$}}
\label{sec:03:e}
Because the Tolman test concerns surface brightness, it is necessary
to cast the S{\'e}rsic profiles in Tables~\ref{tab:03} \& \ref{tab:04} and      
Figure~\ref{fig:03} in terms of surface brightness. The relative shape
of the SB profiles in the S{\'e}rsic family is the subject of this section.    
 
     Once the growth curve of a particular intensity profile is known,
the surface brightness averaged over a given radius follows from,
\begin{equation}
   \langle SB(\eta)\rangle=2.5\log(\pi r_{\eta})^{2} + m(\eta)_{\rm GC}, % (4) 
\label{eq:04}
\end{equation}
where the first term on the right is the area expressed in mag per
square arc sec, and the other is the magnitude from the growth curve
(GC) at the particular value of $\eta$. In practice, if we want only the
run of $\langle$SB$\rangle$ with $\eta$, normalized to say $\eta=2$,
it is sufficient to use  
\begin{equation}
   \Delta\langle SB\rangle_{\eta=2}=5\log r/r(2)+m(\eta)_{\rm GC}.  % (5) 
\label{eq:05}
\end{equation}

     Mean $\langle$SB$\rangle$ relative to that at $\eta=2$ for the
S{\'e}rsic family using equation~(\ref{eq:05}) are listed in
Table~\ref{tab:06} and shown in Figure~\ref{fig:05} for $\eta$ values
from 1.0 to 4.0. The $\langle$SB$\rangle$ values for E galaxies,
calculated from the observed $T=-5$ growth curve, are also listed in
column~(5). Figure~\ref{fig:05} is the second diagnostic diagram that
will be used in later sections. 

% ******************************************************************
%  Figure 5: Correlation of <SB> with Sersic n values
% ******************************************************************

% ******************************************************************
%  Table 6: Surface Brightness Normalized to eta=2 for Eight
%           Sersic Profiles, Averaged Over the Area Interior to the 
%           Listed Eta Radii. The Unit is Magnitudes (Relative to
%           eta=2 Values) Per Unit Area. 
% ******************************************************************

     The close agreement of the listed $\langle$SB$\rangle$ for $T=-5$
with the $n=0.25$ de~Vaucouleurs profile is one of the proofs we are
seeking that S{\'e}rsic profiles have relevance for E galaxies. A
generalization of the proof for different absolute magnitudes for E
galaxies is in the next section.

% ******************************************************************
%     4. THE VARIATION OF THE SURFACE BRIGHTNESS WITH INTRINSIC RADIUS 
%        AT VARIOUS ETA VALUES FOR LOCAL GALAXIES IN THE VIRGO, FORNAX, 
%        AND COMA CLUSTERS
% ******************************************************************
\section{\MakeUppercase{%
         The Variation of the Surface Brightness With Intrinsic Radius 
         at Various} \boldmath{$\eta$} \MakeUppercase{Values for Local
         Galaxies in the Virgo, Fornax, and Coma Clusters}}
\label{sec:04}
The relevance of the S{\'e}rsic family of profiles for observations of
real E galaxies was shown in the last section, but only for the case
of galaxies with the $T=-5$ growth curve. The E galaxies used to
define the $T=-5$ growth curve (Table~11 of the \citeauthor{RC3}) are
those of the average absolute magnitude for such galaxies that exist
in the catalogs of aperture photometry that were available when the
\citeauthor{RC3} was compiled. However, because the luminosity profile
of E galaxies, and hence the S{\'e}rsic $n$ value, is such a strong
function of linear size, we must generalize the \citeauthor{RC3}
$T=-5$ growth curve over a large range of absolute magnitude and
intrinsic size.   

     Data on the observed profiles for local galaxies that span a
range of five magnitudes are in the literature for the Virgo, Fornax,
and Coma clusters. The photometry is by \citet{Fraser:77},  
\citet{King:78}, \citet{Hodge:78}, \citet{Ichikawa:etal:86}, 
\citet{Schombert:86}, \citet{Jedrzejewski:87}, and
\citet{Vigroux:etal:88} for Virgo, 
by \citet{Schombert:86} and \citet{Caldwell:87} for Fornax, and by 
\citet{Oemler:76} for Coma. $\eta$ was calculated in
\citeauthor{SP:90b} from these data using equation~(\ref{eq:02}). 
A diagram similar to Figure~5 in \citeauthor{SP:90a} was drawn for
each galaxy showing the published profile, the growth curve, and
$\eta$ calculated at the listed radii. The $\log r\;$(arc sec) for
$\eta$ values of 1.0, 1.3, 1.5, 1.7, 2.0, 2.5, 3.0, 3.5, and 4.0 were
determined graphically from these diagrams. The surface brightness
averaged over the area interior to the listed radii is listed in
Tables 1-3 in \citeauthor{SP:90b} at each $\eta$ value.  

     Data for 153 cluster galaxies were determined in this way; 83 in
Virgo, 20 in Fornax, and 42 in Coma. The data cover a range in
absolute magnitude from $M_{B}(T) = -23$ to $-13$ and a range of
intrinsic radii from $\log R(\eta=2)$ of 3.3 to 5.1 (pc), and
are listed in Tables~\ref{tab:01}--\ref{tab:03} in
\citeauthor{SP:90b}. Where needed, we adopted distance moduli of 31.7
for Virgo, 31.9 for Fornax, and 35.5 for Coma.

     Calculations were made from these data for the
$\log r(\eta)/r(2)$ radius ratios and the 
$\mbox{SB}(\eta)-\mbox{SB}(\eta=2)$ surface brightness differences
relative to the values at $\eta=2$ over the range of $\eta$ values
from 1.0 to 4.0. These data permit the observations to be compared
with the expectations in the diagnostics Figures~\ref{fig:03} \&
\ref{fig:05} using the family of S{\'e}rsic profiles.   

     Based on the strong correlation of the S{\'e}rsic $n$ exponent with
absolute magnitude discovered by  
\citet[][their Fig.~2]{Binggeli:Jerjen:98}, there must be a
correlation of $\log r(\eta)/r(2)$ with intrinsic radius if the S{\'e}rsic
theoretical profiles are good fits to the observed profiles. This
expectation for the Virgo, Fornax, and Coma cluster galaxies is
verified in Figures~\ref{fig:06} \& \ref{fig:07}
where both $\log r(1)/r(2)$ and
$\langle\mbox{SB}(2)\rangle-\langle\mbox{SB}(1)\rangle$ are plotted
against $\log R(\eta=2)$. Because these ratios change systematically
with $\log R(2)$, and because they are both functions of S{\'e}rsic $n$ at
any given $\eta$ (Figs.~\ref{fig:03} \& \ref{fig:05} here), $n$ must
itself be a function of $\log R(2)$, and therefore of absolute
magnitude.
Figure~\ref{fig:02} of \citet{Binggeli:Jerjen:98} is explicit, showing
that the S{\'e}rsic $n$ varies between 0.15 to 1.5 as the apparent $B_{T}$
magnitude in Virgo cluster E galaxies change between 9 and 18, or
$M_{B(T)}$ between $-22.7$ and $-13.7$.  

% ******************************************************************
%  Figure 6: Variation of the log r(eta=1)/r(2) radius ratio with 
%       intrinsic diameter in pc for the Virgo, Fornax, and Coma cluster
% ******************************************************************
%  Figure 7: Variation of the SB differences with intrinsic diameter
%            in pc for the Virgo, Fornax, and Coma cluster
% ******************************************************************

Interpolating in Table~\ref{tab:04} by using Figure~\ref{fig:06} for
$\log r(1)/r(2)=-0.5$, we determine that the S{\'e}rsic exponent is 
$0.46\pm0.02$ at $\log R(2)=3.9$, which is the average for the
\textsl{HST} clusters. This interpolation shows that $n$ increases as
$R(2)$ decreases, reaching the exponential value of $n=1$ near
apparent magnitude $B(T)=16~ (M_{T}=-15.7)$ in the Virgo cluster 
\citep[see also][]{Binggeli:Jerjen:98}, well into the dE morphological
types. 

     We have also tested the adequacy of the S{\'e}rsic profiles using the
much larger sample of 178 galaxies in 78 local Abell clusters studied
by \citeauthor{SP:91}. The data are in the $V$ band for this
independent sample. They were analyzed in the way just described for
the Virgo/Fornax/Coma clusters. The observed profiles are from 
\citet{Oemler:76}, \citet{Thuan:Romanishin:81},
\citet{Malumuth:Kirshner:85}, and \citet{Schombert:87}. Here, the
intrinsic radius at $\eta=2$ varies from $\log R(2) = 4.0$
to 5.5. From \S~\ref{sec:06} later, we see that this range is larger
than the average $\langle\log R(2)\rangle = 3.9$ for the three remote
\textsl{HST} clusters, and the difference must be accounted for, as is
done in \S~\ref{sec:05} and \ref{sec:06}.  

     A third independent sample of local cluster galaxies is from
\citeauthor{PL:95} who again used first ranked E galaxies in local
Abell clusters. There are 128 galaxies in the PL sample whose
intrinsic radii are again large, ranging from $\log R(2)$ of 4.0 to
5.5.  

     Using these data from the three independent sources, we drew 
diagrams (not shown) similar to Figures~\ref{fig:06} \& \ref{fig:07}
over the additional range of $\eta$ values from 1.3, 1.5, to
1.7. Comparing the variation of $\log r(\eta)/r(2)$ and
$\mbox{SB}(\eta)- \mbox{SB}(2)$ with Figures~\ref{fig:03} \&
\ref{fig:05} for the S{\'e}rsic family as the $n$ exponent is varied shows
that the S{\'e}rsic models are excellent fits to actual E galaxy profiles
over the entire range of $\eta$ and S{\'e}rsic $n$ values, consistent with
Figure~2 of \citet{Binggeli:Jerjen:98} and with the conclusion of
\citet{Kormendy:etal:09}.  

     This completes the proofs that the S{\'e}rsic family of profiles fit
the observed E galaxy data over the total range of E galaxy absolute
magnitudes, and that the curves in Figures~\ref{fig:03} \&
\ref{fig:05} for the S{\'e}rsic family can be used to model the profiles
of real galaxies. The systematically varying $n$ values can be
determined by fitting the data to these two diagnostic diagrams, as we
do in Figure~\ref{fig:15} later.

% ******************************************************************
%     5. CALIBRATION OF THE RELATION BETWEEN SURFACE BRIGHTNESS AND 
%        INTRINSIC RADIUS USING LOCAL E GALAXIES
% ******************************************************************
\section{\MakeUppercase{%
         Calibration of the Relation Between Surface Brightness and 
         Intrinsic Radius Using Local E Galaxies}}
\label{sec:05}
That the surface brightness of E galaxies varies strongly and
systematically with intrinsic radius, becoming brighter as the radius
decreases, has variously been rediscovered in the many papers cited
earlier, perhaps earliest seen in photographs by 
\citet*[][BBC]{BBC:64}. The sense of fainter $\langle$SB$\rangle$ for
larger $R(2)$ values of E galaxies, seen so well in the
\citeauthor{BBC:64} photographs, holds until $\log R(2)=3.8$, where
the relation reverses (\citealt*{BST:84}; \citeauthor{SP:90b},
Fig.~14) for the faint dE galaxies.   

     We showed in \citeauthor{SL:01} that the variation of SB with
intrinsic radius for E galaxies is also well defined at the $\eta$
values of 1.0, 1.3, 1.5, 1.7, and 2.0. The spread at a given radius is
one magnitude whereas the total range of SB between $\log R(2)$ of 4.2
and 5.4 is 4 magnitudes (Fig.~2 of \citeauthor{SL:01}). Because the
Tolman test is about surface brightness, we must either calibrate out
this variation, or we must make the test by comparing the SB of the 
data set at the same $R(\eta)$ radii. We chose the latter here. 

     The analysis in \citeauthor{SL:01} provides the first step in
determining the zero point of the SB-intrinsic radius relation for
local galaxies. The data listed in Table~1 of that reference are
$\eta$, $\langle$SB$\rangle$, and intrinsic size for 118 first ranked
local Abell clusters. These have been calculated from
equations~(\ref{eq:01}), (\ref{eq:04}), and (\ref{eq:05}) using the
observed growth curves of \citeauthor{PL:95} with $H_{0}=50$. 

     The $R$ magnitudes by \citeauthor*{PL:95} are on the Cape/Cousins
near red photometric system as realized by
\citet{Landolt:83,Landolt:92}. The $R_{\rm cape}$ band pass differs
from $R_{J}$ of \citet{Johnson:65}, which is identical to the
$r(\mbox{S20})$ system of \citet{Sandage:Smith:63} that I used for all
the Mount Wilson/Palomar near-red photometry of first ranked cluster
galaxies with S20 photocathods in the 1970s
\citep[eg.][]{Sandage:72,Sandage:73}. That photometric $R$ system has
now been largely replaced in the literature by the Cape/Cousins
$(RI)_{\rm cape}$ system because of the work of Landolt with his
all-sky $UBV(RI)_{\rm cape}$ standards. The difference in zero point
between the $R_{\rm cape}$ and $R_{J}$ systems is $0.26\mag$ at the
color of local E galaxies, with $R_{\rm cape}$ being fainter at this
color \citep{Sandage:97}.  
The $\langle$SB$\rangle_{R_{\rm cape}}$ and $\log R(\eta)$ data for
the PL sample are in Table~1 of \citeauthor{SL:01}. 
 
     Least squares linear fits to the $\langle$SB$\rangle$, $\log R$
data of \citeauthor*{PL:95} are in Table~2 of \citeauthor{SL:01} and
shown in Figure~\ref{fig:02} there for $\eta$ values of 1.3, 1.5, 1.7,
and 2.0. These linear fits are valid only for $\log R > 4.0$. The
$\langle$SB$\rangle$ data have been reduced to zero redshift by making
them brighter by $0.16\mag$ than the values measured by
\citeauthor{PL:95}, determined by the method described in
\citeauthor{SL:01}.   

     The linear correlation of $\langle$SB$\rangle$ and size for
$\eta=2.0$ for zero redshift, is, 
\begin{equation}
    \langle\mbox{SB}(2)\rangle_{R_{\rm cape}}=2.97\log R(2)_{50}+8.53, % (6)
\label{eq:06}
\end{equation}
valid for $\log R(2)$ (pc) $> 4.3 (H_{0}=50)$. The redshifts are small 
enough that the value of the spatial curvature, $q_{0}$, is 
inconsequential. 

     The extension of equation (\ref{eq:06}) to $\log R(2)<4.3$ is
nonlinear, bending toward fainter $\langle$SB$\rangle$ at the smaller
radii, shown in Figures~1 \& 2 of \citeauthor{SL:01}. It was accounted
for there by a nonlinear addition to equation~(\ref{eq:06}) for $\log
R(2)<4.4$. A first approximation for the correction was made from
Table~3 in \citeauthor{SL:01}. 
It is revised in Table~\ref{tab:07} here. The correction is smaller
than in \citeauthor{SL:01} by $0.3\mag$ at $\log R(2)=3.4$, the
difference decreasing to zero at $\log R(2)=4.4$. The new corrections
here are derived from the combined $\langle$SB$\rangle$, $\log R$
correlations in the Virgo, Fornax, and Coma clusters, discussed above,
by a better spline connection of the cluster data from
\citeauthor{SP:90b} above and below $R(2)=4.4$ (pc).  

% ******************************************************************
%  Table 7: Nonliner Corrections to Equations (6)-(9)
% ******************************************************************

     A second calibration, independent of equation~(\ref{eq:06}), is
made using the extensive $V$ band data from Table~1 of
\citeauthor{SP:90b}, discussed earlier. The least squares regression
of SB$_{V}$ on $\log R(2)_{50}$ for the 178 galaxies in this earlier
sample is,
\begin{equation}
            \mbox{SB}_{V}(2) = 2.89 \log R(2) + 9.53,        % (7)
\label{eq:07}
\end{equation}
reduced to zero redshift and valid for $\log R > 4.3$. The
nonlinearity for $\log R < 4.3\ (H_{0}=50)$ is clearly seen (not  
shown) in these data, needing again the Table~\ref{tab:07} correction
for the smaller radii. 

     To compare (\ref{eq:07}) with equation~(\ref{eq:06}) we change
the slope to 2.97 and adjust the zero point, giving, 
\begin{equation}
            \mbox{SB}_{V}(2) = 2.97 \log R(2) + 9.17.        % (8). 
\label{eq:08}
\end{equation}
again valid only for $\log R(2) > 4.3 (H_{0}=50)$. With these changes,
equation~(\ref{eq:08}) is everywhere within $0.08\mag$ of equation
(\ref{eq:07}) over the range of $\log R$ between 4.0 and 5.0. 

     To compare equation~(\ref{eq:08}) in $V$ with
equation~(\ref{eq:06}) in $R$ we need the color index, 
$(V\!-\!R)_{\rm cape}$, for E galaxies with linear sizes of $\log
R(2)$ between 4.0 and 5.0. The $(V\!-\!R)$ color for such giant E
galaxies have been determined from the catalog of growth curves in $R$
and $I$ by \citet{deVaucouleurs:Longo:88} for a variety of photometric
systems. Choosing from those that are on the Cape/Cousins system
(listed as $R'$ and $I'$ in their catalog) for a sample of 17 local
$T=-5$ galaxies with 66 measurements gave the mean color indices of 
$(V\!-\!R)_{\rm cape}=0.57\pm0.003$, and $(R\!-\!I)_{\rm
  cape}=0.64\pm0.003$. These are nearly identical with the values by  
\citet{Poulain:Nieto:94} from a larger sample. 

     Applying $V-R = 0.57$ to equation~(\ref{eq:08}) gives,
\begin{equation}
          \mbox{SB}(2)_{R_{\rm cape}} = 2.97 \log R(2) + 8.60,     % (9)
\label{eq:09}
\end{equation}
for local E galaxies with $\log R(2) > 4.3~(H_{0}=50)$. This is within
$0.08\mag$ of the $\langle$SB$\rangle$, $\log R(2)$ calibration in
equation~(\ref{eq:06}) which is based on the independent photometry of
\citeauthor*{PL:95}. In what follows we adopt equation~(\ref{eq:09}),
and correct it by Table~\ref{tab:07} for smaller radii. 

     Table~\ref{tab:08}, based on equation~(\ref{eq:09}), corrected by
Table~\ref{tab:07}, is our final adopted calibration for local zero
redshift E galaxies for the nonlinear relation between $\log R(2)$ and
$\langle$SB$\rangle$. The entries are calculated from the
$\langle$SB$\rangle_{R}$ values just described using
$(V-R)_{\rm cape}=0.57$ and $(R-I)_{\rm cape}=0.64$. 

% ******************************************************************
%  Table 8: Ridge-Line Variation of <SB> with the log R Intrinsic Size
%           at eta=2 at Zero Redshift for E Galaxies in the
%           V(RI)_Cape Pass Bands
% ******************************************************************

% ******************************************************************
%     6. SURFACE BRIGHTNESS PROFILES FOR THE THREE HST CLUSTER 
%        GALAXIES COMPARED WITH THE LOCAL SAMPLE: THE TOLMAN TEST
% ******************************************************************
\section{\MakeUppercase{%
         Surface Brightness Profiles for the Three \textsl{HST} Cluster 
         Galaxies Compared with the Local Sample: the Tolman Test}}
\label{sec:06}
Surface brightness magnitudes averaged over $R(\eta)$, the $R(\eta)$
sizes, and absolute magnitudes are listed in Tables~2--4 of  
\citeauthor{LS:01c} for 34 galaxies in the three \textsl{HST} clusters. 
The data in \citeauthor{LS:01b} show that the accuracies of $\eta$ and
$\langle$SB$\rangle$ are better than 2\% at radii between 0.1 and 1
arc~sec. The intrinsic $R$ sizes for the database range from $\log
R(2)= 3.45$ to 4.20 $(H_{0}=50, q_{0}=1/2, \Lambda = 0)$, with a mean
of 3.9.

     Because of the strong variation of $\langle$SB$\rangle$ with
intrinsic size, which we have been emphasizing, we have divided the
galaxies in the \textsl{HST} clusters into three bins of different
intrinsic radii, and make the Tolman test in each bin separately to
compensate for the variation.                    

     Table~\ref{tab:09} lists the $\langle$SB$\rangle$, 
$\log R(\eta)$ data at the $\eta$ values of 1.0, 1.3, 1.5, 1.7, and
2.0 for the three radius bins in \textsl{HST} Cl 1324+3011. The root
data are from Table~3 of \citeauthor{LS:01c}. The bottom entries are
the mean values.

     To make the Tolman test we need the $\langle$SB$\rangle$, 
$\log R(\eta)$ relation for galaxies of zero redshift at the mean 
$\log R(2)$ for the three \textsl{HST} radius bins in
Table~\ref{tab:09}. An example, reading from Table~\ref{tab:09}, is
that the mean radius at $\eta=2$ for the largest radius group in 
Cl 1324+3011 is $\log R(2)= 4.087$ (in parsecs).  

% ******************************************************************
%  Table 9: HST <SB> Data For a Range of eta Values for Cl 1324+3011
%           Binned Into Three Radius Groups.  
% ******************************************************************
%  Table 10: Zero Redshift Surface Brightness - eta Relation For a
%            Sersic Profile With n=0.46 at the Mean log R(2) for the
%            Three Radius Groups for HST Cl 1324+3011  
% ******************************************************************

     The $\langle$SB$\rangle_{I}$-radius relation at $\eta=2.0$ and
$\log R(2)=4.087$ for local galaxies is interpolated for the standards
in Table~\ref{tab:08}. The surface brightness of
$\langle$SB$_{I}(2)\rangle=22.86\;$mag/arcsec$^2$ at $\eta=2.0$ is
then spread to the $\eta$ values of 1.0, 1.3, 1.5, and 1.7 using the
surface brightness ratios in the $n=0.46$ column of
Table~\ref{tab:06}. This procedure gives the standard
$\langle$SB$\rangle, \eta$ curve for zero redshift at the five
fiducial $\eta$ values for local E galaxies. 

     The result for Cl 1324+3011 ($z=0.7565$) is shown in 
Figure~\ref{fig:08}. The zero-redshift $\langle$SB$\rangle_{I}$,
$\eta$ curves from Table~\ref{tab:10}, valid at the marked mean
$R(\eta=2)$ radii for each group, are shown in the upper part of the
diagram. The data from Table~\ref{tab:09} are plotted in the lower
part for the three radius bins. The smooth curves are interpolated
between the points. The bin number is marked at the left. Clearly, a
Tolman signal is present. The family of observed curves is fainter
than the family of zero-redshift calibration curves.  

% ******************************************************************
%  Figure 8: SB vs. eta for Cl 1324+3011
% ******************************************************************

     To see this signal more clearly, the data in each radius bin 
are plotted separately in Figure~\ref{fig:09}. Error bars are put on
the observed points. Zero-redshift standard curves from
Table~\ref{tab:10}, are shown as dashed. The shape of the S{\'e}rsic
curve for $n=0.46$ is brought down from the dashed curve near the top
of each panel in Figure~\ref{fig:09} and is zero-pointed to the data
at $\eta=1.5$.  

% ******************************************************************
%  Figure 9: data from Figure 8 plotted separately (Cl 1324+3011)
% ******************************************************************

     The fit of \textsl{HST} data to the standard curve is
excellent. The $\langle$SB$\rangle$ difference between the
zero-redshift standard (upper curves) and the observed points is the
effect we are seeking. It is the Tolman signal as modified by the
luminosity change in the remote 1324 + 3011 cluster due to evolution
in the look-back time.

     The same analysis is made in Figures~\ref{fig:10} \& \ref{fig:11}
for Cl 1604+4304 ($z=0.8967$) from the data in Tables~\ref{tab:11}
\& \ref{tab:12} (there are only two radius bins), and in 
Figures~\ref{fig:12} \& \ref{fig:13} for Cl 1604+4321  
($z=0.9243$) from Tables~\ref{tab:13} \& \ref{tab:14}.             

% ******************************************************************
%  Table 11: Same as Table 9 for HST Cl 1604+4304. 
% ******************************************************************
%  Table 12: Standard eta-Surface Brightness Curves at Zero Redshift
%            for the Two Radius Bins for HST Cl 1604+4304 and for
%            Four Arbitrary Bins for Illustration. 
% ******************************************************************
%  Table 13: Same as Tables 9 & 11 but for HST Cl 1604+4321. 
% ******************************************************************
%  Table 14: Same as Tables 10 & 12 but for HST Cl 1604+4321. 
% ******************************************************************

% ******************************************************************
%  Figure 10: SB vs. eta for Cl 1604+4304
% ******************************************************************
%  Figure 11: data from Figure 10 plotted separately (Cl 1604+4304)
% ******************************************************************
%  Figure 12: SB vs. eta for Cl 1604+4321
% ******************************************************************
%  Figure 13: data from Figure 12 plotted separately (Cl 1604+4321)
% ******************************************************************

     The surface brightness differences between the remote and the
local galaxies, read from Figures~\ref{fig:09}, \ref{fig:11}, and
\ref{fig:13}, are collected in Table~\ref{tab:15}. The entries can be
compared with Tables~5, 6, \& 7 of \citeauthor{LS:01c}. They are
similar but not identical. The small difference between the studies is
because the method of search here for the Tolman signal is not the
same. There we accounted for the change of $\langle$SB$\rangle$ with
intrinsic size by comparing the zero points of the
$\langle$SB$\rangle$, radius curve at {\em different\/} $\eta$ values
(Figs.~1 and 2 there).  
Here we compare the local and remote data at {\em fixed radii\/} as
$\eta$ is varied (Figs.~\ref{fig:08}--\ref{fig:12} here). Part of the
difference is also due to the slight change in the zero redshift
standard $\langle$SB$\rangle$-$\log R$ relation in
Table~\ref{fig:08}. Each method uses the same observational data and
with each we reach the same conclusion, which is this.           

% ******************************************************************
%  Table 15: Difference in the Surface Brightness Between the HST
%            Clusters and the Zero Redshift Standard SB vs. eta Curves
%            for Each of the Radius Bins at Each of the Fiducial eta
%            Values. 
% ******************************************************************

     The differences in $\langle$SB$\rangle$ are large between remote
and local galaxies over parts of E galaxy images defined by Petrosian
$\eta$ radii, but in each case are smaller than the $(1+z)^{4}$ Tolman 
prediction. We interpret this to mean that a Tolman $(1+z)^{4}$
cosmological signal exists but is degraded by luminosity evolution
that amounts to approximately $0.8\mag$ in $R$ and $0.4\mag$ in
$I$. The exact values at the bottom of Table~\ref{tab:15} are listed
for each of the five fiducial $\eta$ positions in the mean of the E
galaxy images for each radius bin.
     
% ******************************************************************
%  6.1. The Estimated Errors
% ******************************************************************
\subsection{The Estimated Errors}
\label{sec:06:a}
Before the error budget can be determined, two caveats are necessary. 

     (1) In Figures~\ref{fig:08}--\ref{fig:13} we have used an
$n=0.46$ S{\'e}rsic exponent to define the shape of the zero redshift
$\langle$SB$\rangle$-$\eta$ relation in every radius bin for the mean
radius in each. However, following \citet[][their
Fig.~2]{Binggeli:Jerjen:98}, we showed in Figures~\ref{fig:06} \&
\ref{fig:07} using Tables~\ref{fig:04} and \ref{fig:06} that the
S{\'e}rsic $n$ exponent is a strong function of the intrinsic $R(2)$
size. Hence, we should have used {\em different\/} standard
zero-redshift curves with different $n$ values that are relevant for
the mean radius for each of the particular bins. We have ignored this
detail because it makes a negligible difference in Table~\ref{tab:15}
and Figures~\ref{fig:08}--\ref{fig:13}.  

     Nevertheless it is of interest to estimate the variation of
effective $n$ values for the \textsl{HST} cluster galaxies. 
Figure~\ref{fig:14} with Table~\ref{tab:04} is useful in making the
estimate. These show that, as the size varies between $\log R(2)=3.5$
and 4.2, the S{\'e}rsic exponent varies between 0.4 and 0.6 for the 
\textsl{HST} galaxies. The mean over all the \textsl{HST} galaxies is
$n=0.46$, shown in Figure~\ref{fig:15} from the data listed in 
Table~\ref{tab:16}.  

% ******************************************************************
%  Table 16: Average log r(eta)/r(2) vs. eta for the HST Clusters + Average 
% ******************************************************************

% ******************************************************************
%  Figure 14: log R(1)/R(2) vs. log R(2) for the HST cluster galaxies
% ******************************************************************
%  Figure 15: log r(eta)/r(2) vs. eta for the mean HST clusters
% ******************************************************************

     (2) The second caveat is this. We have calculated the intrinsic
sizes using a world model with $q_{0}=1/2$, $H_{0}=50$,
$\Lambda=0$. Different radii would be obtained for different $q_{0}$
models. Table~8 of \citeauthor{LS:01c} shows that if $q_{0}=0$, the
radii would be about 25\% larger than we have used. If $q_{0}=1$, the
$R$ values would be 15\% smaller. Therefore, using the slope of
equation~(\ref{eq:09}), the surface brightness of the standard zero
redshift curves would be displaced relative to the \textsl{HST} points
by $0.3\mag$ brighter and $0.18\mag$ fainter from the offsets we use
here for $q_{0}=1/2$.

     We have not calculated the effect on $R$ using a finite value of
$\Lambda$, taken from the current ``concordance world model'', with
assigned values of $\Lambda$ and $q_{0}$ because the uncertainties
here are at the level just stated for the range of $q_{0}$ from 0 to
1, and are bracketed by them.

     This amounts to a {\em systematic\/} error over which we have no
control unless we know the correct world model, which we do not. We
ignore this systematic uncertainty in what follows, giving only the
statistical errors for the $q_{0}=1/2$ case.  

     Ignoring these two caveats, the statistical errors are estimated
as follows. We require the mean $\langle$SB$\rangle$ offsets in the  
$\langle$SB$\rangle$ vs.\ $\eta$ curves between the zero redshift
standard curves and the data points in Figures~\ref{fig:09},
\ref{fig:11}, and \ref{fig:13}. There are two components. (a) One is
the difference between the observed points  and the standard curves
brought down from the zero redshift $\langle$SB$\rangle$ curves. (b)
The second is the accuracy with which we know the position of the
standard curves at zero redshift for particular radius values.     

     (a). Table~\ref{tab:15} lists the individual $\langle$SB$\rangle$
offsets for each fiducial $\eta$ Petrosian homology position in each
of the radius bins. The rms differences from the standard curves are
small, averaging $0.06\mag$ for each cluster in each radius bin. In
addition, there is no correlation of the residues with $\eta$, showing
that the data have the same shape, on average, as the standard 
$n=0.46$ S{\'e}rsic profile. The mean rms of the $\langle$SB$\rangle$
differences with the standard curve averages $0.12\mag$ over the range
of $\eta$ from 1.0 to 2.0. With five data points in each group, the
average error of the offset is $\pm0.06\mag$.

     (b). As read from the error in the least squares intercept in
Table~2 of \citeauthor{SL:01}, the accuracy with which we know the
placement of the zero-redshift standard $\langle$SB$\rangle$-$\eta$
curve {\em at fixed radius\/} is $\pm0.08\mag$. However, the radius in
each size bin itself varies within the bin, and its mean radius has an
rms variation that also introduces a contribution to the placement
error. These variations in $\langle R\rangle$ are listed within the
body of Tables~\ref{tab:09}, \ref{tab:11}, and \ref{tab:13}. They
average $\mbox{rms}=0.100$, giving a mean error of $\pm0.05$ in 
$\log R$. From equation~(\ref{eq:09}) this translates to an mean error
in the placement of the standard SB curves at zero redshift of
$\pm0.15\mag$. Adding this in quadrature to $\pm0.08\mag$ gives a
total mean placement error of the zero-redshift curves of
$\pm0.17\mag$. Adding this in quadrature to the $\pm0.06\mag$ mean
error of the \textsl{HST} data points gives a total uncertainty of the
mean $\langle$SB$\rangle$ offsets in Table~\ref{tab:15} as $\pm0.18\mag$. 

     From Table~\ref{tab:15} we adopt the mean $\langle$SB$\rangle$
offsets between the \textsl{HST} galaxies and the local E galaxies of
the same size as $2.04\pm0.18\mag$ for Cl 1324+3011 ($z=0.7565$) in
the rest frame $I$ band, $2.51\pm0.18\mag$ for Cl 1604+4304
($z=0.8967$) in the rest frame $I$ band, and $1.99\pm0.18\mag$ for 
Cl 1604+4321 ($z=0.9243$) in the rest frame $R$ band. 

     These offsets correspond to Tolman plus luminosity evolution
signals of 
$p=3.33\pm0.30$ for Cl 1324+3011 in the $I$ band, 
$p=3.61\pm0.26$ for Cl 1604+4304, also in the $I$ band, and 
$p=2.80\pm0.25$ for Cl 1604+4321 in the $R$ band, where $p$ is the
exponent on $(1+z)$. Combining the $p$ values for the two clusters in
the $I$ band, and repeating the answer in the $R$ band for 
Cl 1604+4321 gives the final answers as,  
\begin{equation}
   \Delta\langle\mbox{SB}\rangle =\begin{cases}
2.5\log(1+z)^{2.80\pm0.25} & \mbox{mag in the\ } R\ \mbox{band}   \\
2.5\log(1+z)^{3.48\pm0.14} & \mbox{mag in the\ } I\ \mbox{band}.
\end{cases}
\label{eq:10}
\end{equation}

% ******************************************************************
%  6.2. Evolution in the Look-Back Time
% ******************************************************************
\subsection{Evolution in the Look-Back Time}
\label{sec:06:b}
If the true Tolman signal is $(1+z)^{4}$, then the component due to
luminosity evolution is $(1+z)^{4-p}$, which is,       
\begin{equation}
   M_{\rm evol} =\begin{cases}
2.5\log(1+z)^{1.20\pm0.25} & \mbox{mag in the\ } R\ \mbox{band}   \\
2.5\log(1+z)^{0.52\pm0.14} & \mbox{mag in the\ } I\ \mbox{band}.
\end{cases}
\label{eq:11}
\end{equation}
For redshifts of $z=0.86$, which is the mean for the three \textsl{HST}
clusters, these luminosity evolutions are $0.81\mag$ in rest frame $R$
and $0.35\mag$ in rest frame $I$. Such luminosity changes at this mean
redshift are consistent with the stellar evolution models of 
\citet{Bruzual:Charlot:93} if the initial star formation was a burst
near the beginning of the creation of the galaxy (\citeauthor{LS:01c},
\S~4.2). Equations~(\ref{eq:10}) and (\ref{eq:11}) are closely the
same as Lubin and I (\citeauthor{LS:01c}) found, using a different
representation of the \textsl{HST} data.

% ******************************************************************
%  Q.E.D.
% ******************************************************************
\subsection*{Q.E.D.}
I have been told by my son John, who majored in physics at University
of California, Davis that the professor of advanced mechanics would
post the solutions to the weekly five assigned problems a week after
they were due, and always signed the solutions Q.E.D. As the
difficulty of the problems increased week by week, some of the
students grew increasingly disturbed by what they conceived to be a
mocking by the professor, because they thought that Q.E.D. meant
``Quite Easily Done'' rather than ``Quod Erat Demonstrandum'', meaning
``As Was To Be Demonstrated''.  

     The solution of the Tolman test given by \citeauthor{LS:01c} and
differently here has not been quite as easily done as I set out at the
planning meeting for the \textsl{LST} in 1974. It has required many
developments not yet made at the time. However, the test seems to have
been successful. The Tolman prediction is verified. The expansion would
seem to be real.

% ******************************************************************
%     7. SUMMARY AND DISCUSSION
% ******************************************************************
\section{\MakeUppercase{Summary and Discussion}}
\label{sec:07}
%
% ******************************************************************
%  7.1. Preliminaries   
% ******************************************************************
\subsection{Preliminaries}   
\label{sec:07:a}
One of the major problems that hampered progress was how to define a
homologous radius at which to compare the surface brightness of local
and remote E galaxies. An obvious early choice was the half-light
radius $r_{e}$, but this was hard to measure before the advent of area
detectors. To determine $r_{e}$ requires a growth curve that extends
to ``infinite'' radius so that it can be backed off by $0.75\mag$ to
the half-light value. Finding the asymptotic ``total'' magnitude
depends on an assumed luminosity profile which is generally not the
standard de~Vaucouleurs $r^{1/4}$ curve. 

     Another scale-free radius is the Hubble $a$ fitting parameter,
used in my first proposal for the test with an \textsl{LST}. This is
the radius where the measured surface brightness is 1/4 of the central
value. However, this is even harder to measure. It requires knowing
the central intensity, which is elusive because of insufficient
spatial resolution of the telescope, the detector, and from the
ground, the seeing.    

     The solution has been to use Petrosian radii, defined as the
comparison of the SB at a particular radius to the average  
surface brightness, $\langle$SB$\rangle$, inside that radius. 
We have formulated the Tolman test, both in 
\citeauthor{SP:90a},b, \citeauthor{LS:01a},b,c
and here by using Petrosian radii throughout. 

     The properties of various ratios of measurable parameters at
Petrosian $\eta$ values of 1.0, 1.3, 1.5, 1.7, and 2.0 mag is the
subject of \S~\ref{sec:03}. The two diagnostic diagrams of
Figures~\ref{fig:03} \& \ref{fig:05} are central to the
discussion. They are related to the luminosity profile, permitting the
assigning of a particular S{\'e}rsic profile to the data but also
providing a way to calculate the $\langle$SB$\rangle$ over a range of
$\eta$ radii when the $\langle$SB$\rangle$ is known at $\log R$ at
$\eta=2$. Based on these two diagrams, we have used 11 steps to
complete the Tolman test. 
    
% ******************************************************************
%  7.2. The Eleven Steps
% ******************************************************************
\subsection{The Eleven Steps}
\label{sec:07:b}
Only the shape of the $\langle$SB$\rangle$ vs.\ $\eta$ standard curve
is given by the second diagnostic diagram in Figure~\ref{fig:05}.
Once the S{\'e}rsic $n$ exponent is known, the shape of its standard curve
must be calibrated in zero point. We proceed in eight steps for this 
calibration. 
 
     (1). By the method in \S~\ref{sec:05} we determine the zero point
of the $\langle$SB$\rangle$ vs.\ $\log R$ zero redshift curve at the $R$
radius corresponding to $\eta=2.0$. The results are in 
equations~(\ref{eq:06})--(\ref{eq:09}) in the $V$ and $R_{\rm cape}$
band passes, valid for $\log R > 4.3$.  

     (2). For smaller radii, non-linear corrections to these equations
have been determined from photometric data in the Virgo, Fornax, \&
Coma clusters. These are listed in Table~\ref{tab:07}.    

     (3). Applying Table~\ref{tab:07} to equation~(\ref{eq:09}) gives the
$\langle$SB$\rangle$ averaged over $R$ at zero redshift vs.\ the $\log
R$ values in Table~\ref{tab:08}, tabulated in $V$, $R$, and $I$ for
intrinsic radii at $\eta=2.0$ over the range of $\log R$ from 5.4 to
3.4 (pc). 

     (4). The calibration of $\langle$SB$\rangle$ vs.\ $R$ at $\eta$
values other than 2.0 is found by spreading the calibration of
Table~\ref{tab:07} to the four fiducial $\eta$ values between 2.0 to
1.0 by using the diagnostic diagram of Figure~\ref{fig:05}
(Table~\ref{tab:06}) with a S{\'e}rsic exponent of $n=0.46$, determined as
follows.   

\begin{figure}[t]
   \plotone{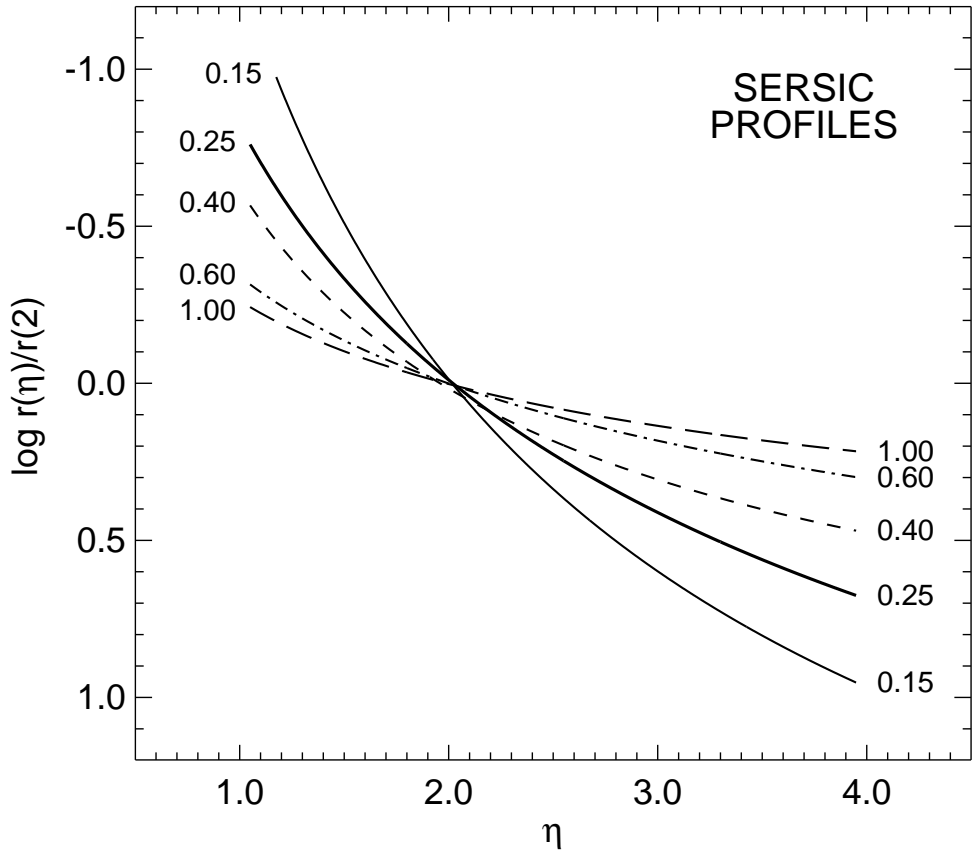}
   \caption{Correlation of $\eta$ with the $\log r(\eta)/r(2)$ radii
            ratio for six S{\'e}rsic profiles from the data in 
            Table~\ref{tab:04}.}
\label{fig:03}
\end{figure}

\begin{figure}[t]
   \plotone{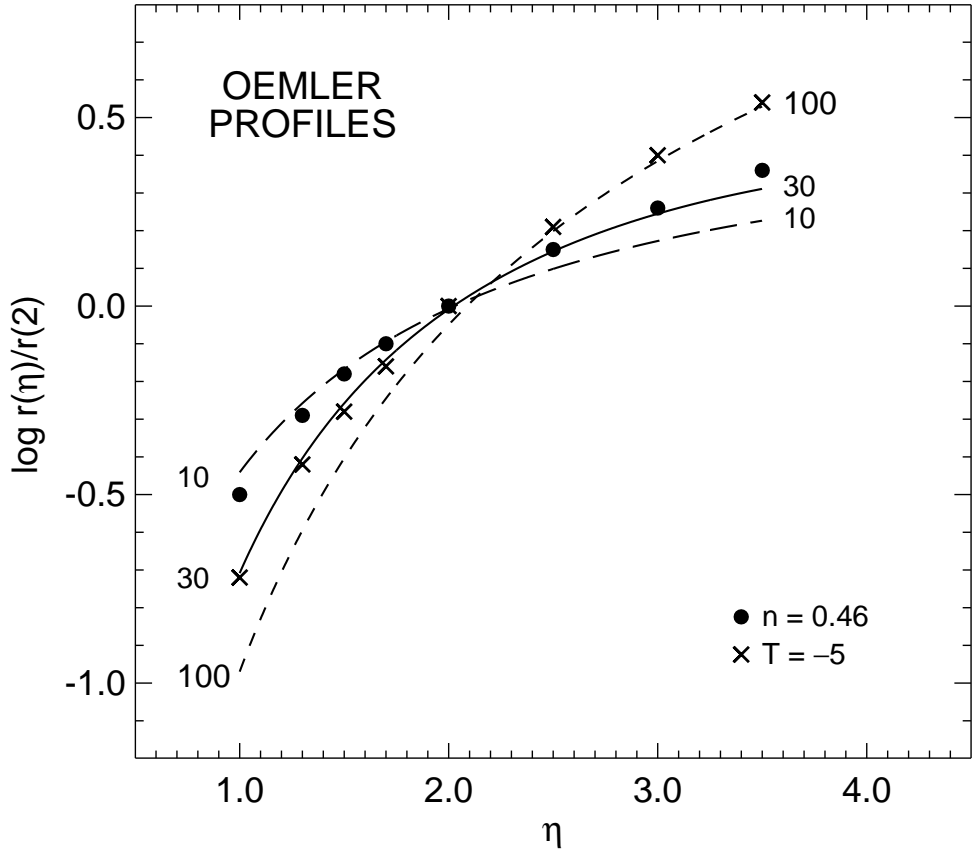}
   \caption{Three Oemler profiles in $\log r/r(2)$ vs.\ $\eta$ compared 
      with the observed $T=-5$ (giant E) galaxy profile and the $n=0.46$ 
      S{\'e}rsic profile. The data are from Tables~\ref{tab:03} and 
      \ref{tab:05}.} 
\label{fig:04}
\end{figure}

\begin{figure}[t]
   \plotone{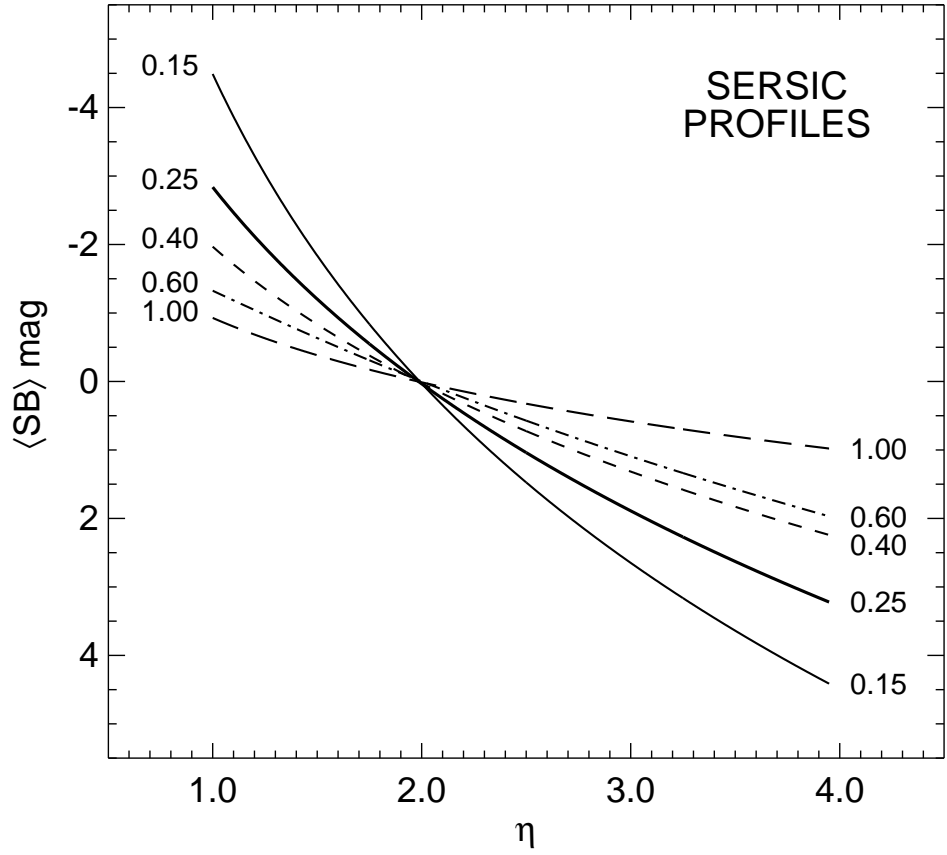}
   \caption{Correlation of $\langle$SB$\rangle$ with S{\'e}rsic $n$ values
      from Table~\ref{tab:06}.}
\label{fig:05}
\end{figure}

     (5). Proof that the S{\'e}rsic family of profiles is appropriate 
and that the correct S{\'e}rsic $n$ exponent can be found is made by 
recovering the discovery, made by the many authors cited, that $n$ 
varies strongly with absolute magnitude. From Figures~\ref{fig:07} and
\ref{fig:08} in \S~\ref{sec:04}, it is shown that both the 
$\log r(\eta=1)/r(2)$ radii ratios in Figure~\ref{fig:03} and the
$\langle\mbox{SB}(2)\rangle-\langle\mbox{SB}(1)\rangle$
differences in Figure~\ref{fig:05} vary systematically with 
$\log R$. That this must be so, provided that the S{\'e}rsic family is a
good fit to E galaxy profiles, follows from the discovery by
\citet{Binggeli:Jerjen:98} that the S{\'e}rsic $n$ exponent varies between
0.15 and 1.5 as the absolute magnitude of E galaxies becomes fainter
between $M_{B}=-23$ and $-13$ for Virgo cluster galaxies. 

\begin{figure}[t]
   \epsscale{0.73} % for Figs. 6+7
   \plotone{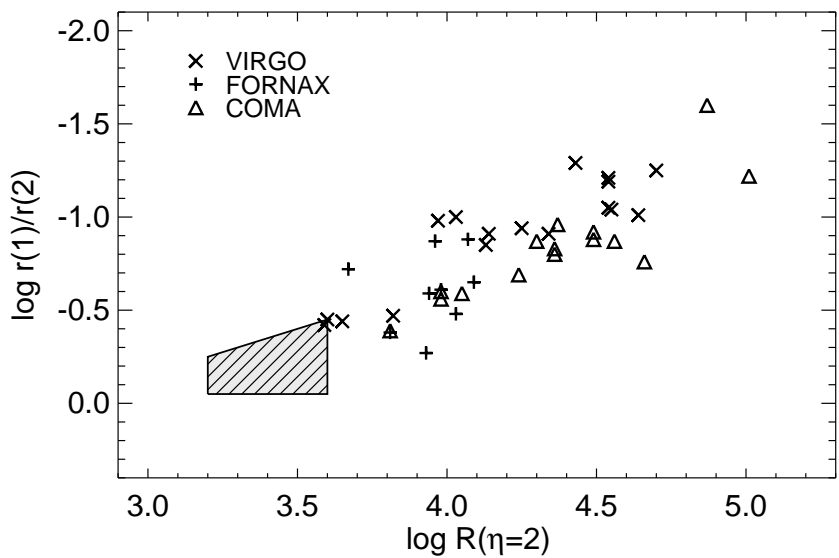}
   \caption{Variation of the $\log r(\eta=1)/r(2)$ radius ratio with 
      intrinsic diameter in pc for the Virgo (skipping jack crosses), 
      Fornax (Roman crosses), and Coma cluster galaxies (triangles) 
      from the data in Tables~1--3 of \citeauthor{SP:90b}. The region
      of the dwarf dE galaxies for $\log R < 3.6$ is hatched. 
      Interpolating in the diagnostic Table~\ref{tab:04}, or using
      Figure~\ref{fig:03}, shows that $n$ for E and dE galaxies varies
      between 0.15 and 1.0 over the range of $\log R(2)$ between
      3.3 and 4.8.}
\label{fig:06}
\end{figure}

\begin{figure}[t]
   \plotone{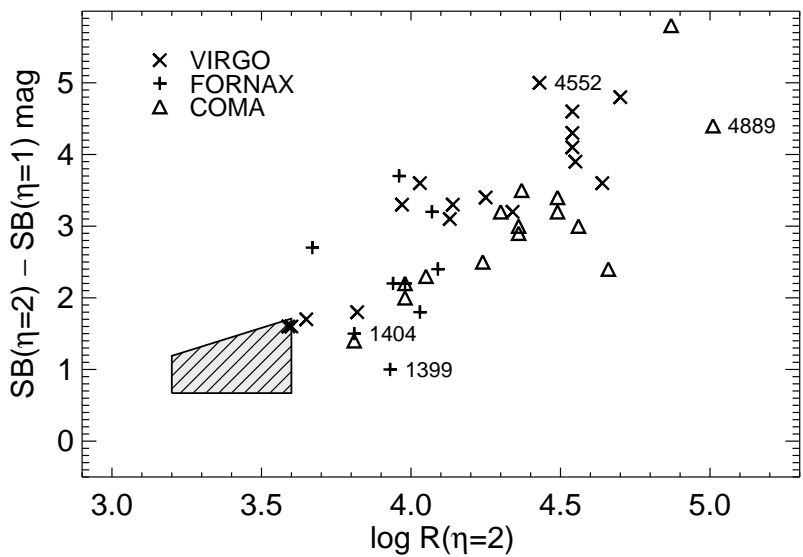}
   \caption{Same as Figure~\ref{fig:06} but for the surface brightness
      differences between $\eta=1$ and 2 for Virgo, Fornax, and
      Coma from the data in Tables~1--3 in \citeauthor{SP:90b}.}
\label{fig:07}
\end{figure}

\begin{figure}[t]
   \epsscale{0.85} % for Figs. 1-5, 8-15
   \plotone{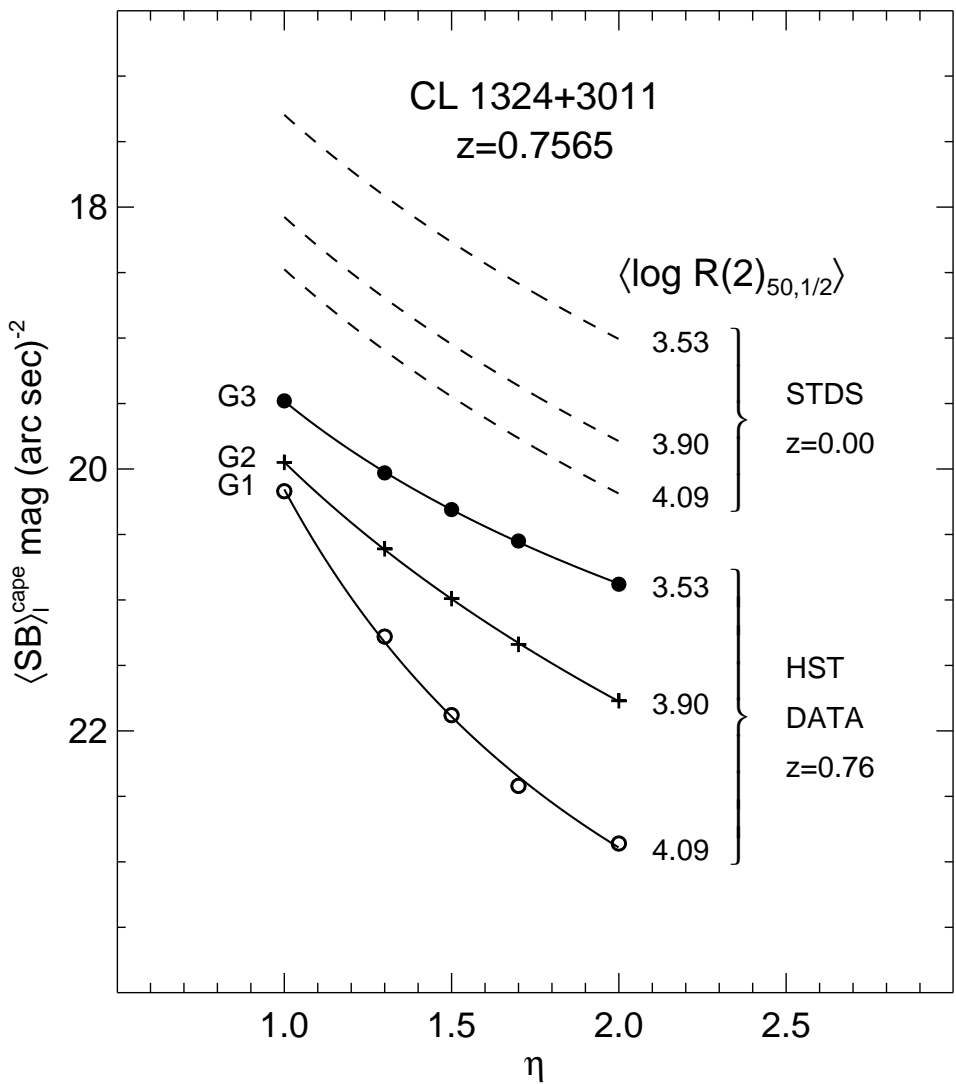}
   \caption{Surface brightness in the Cape/Cousins $I$ band vs.\ $\eta$ for 
      galaxies in Cl 1324+3011 ($z=0.7565$) for three different intrinsic
      radii bins at $\eta=2.0$. Standard zero-redshift curves at the listed
      radii, $R(2)$, for an $n=0.46$ S{\'e}rsic profile are the dashed curves
      near the top, listed in Table~\ref{tab:10}. Smooth curves with the same
      shape are put through the observed data (Table~\ref{tab:09}) in the
      lower part of the diagram. Data are from Tables~\ref{tab:09} and
      \ref{tab:10}.}
\label{fig:08}
\end{figure}

     (6). Reading Figure~\ref{fig:06} at $\log R=3.9$, which is the
average for galaxies in the in the \textsl{HST} sample, and
interpolating with $\langle\log r(1)/r(2)\rangle=-0.50$ between the
S{\'e}rsic $n$ values, gives $n=0.46$. 

     (7). Reading Table~\ref{tab:06} at this $n$ gives the
$\langle$SB$\rangle$ differences with $\langle$SB($\eta=2.0$)$\rangle$
needed to spread the calibration from $\eta=2$ to the other four
fiducial $\eta$ values we are using. 

     (8). This calibration of the $\langle$SB$\rangle$-$\log R$
relation for local E galaxies at zero redshift at these five $\eta$
values at the mean radii of the size bins used for the \textsl{HST}
clusters is listed in Tables~\ref{tab:10}, \ref{tab:12}, \&
\ref{tab:14} of \S~\ref{sec:06}.  

     This completes the calibration steps. The remaining steps to 
the Tolman test itself are three. 

\begin{figure}[t]
   \plotone{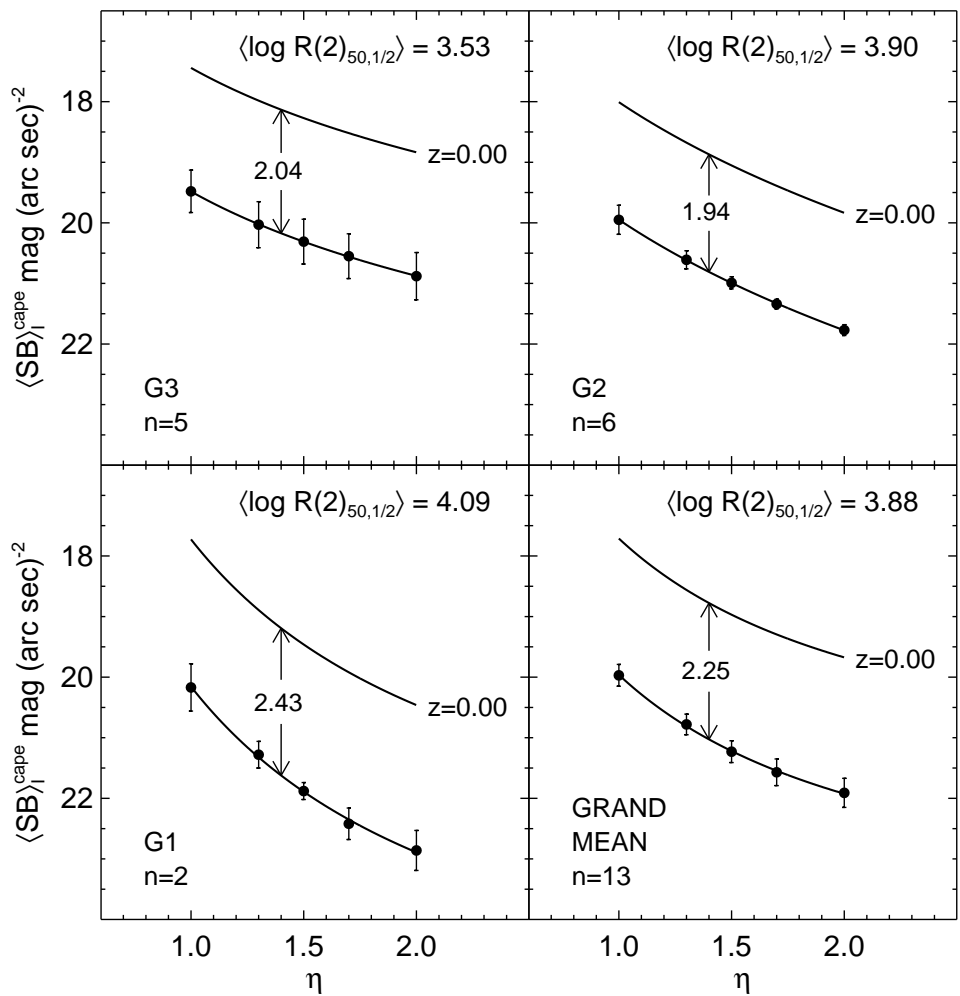}
   \caption{The data from Figure~\ref{fig:08} plotted separately for
      the three radius bins and for the mean for Cl 1324+3011. The shape
      of the standard profile at each mean radius is dropped onto the
      data from the upper curves, zero pointed at $\eta=1.5$.}
\label{fig:09}
\end{figure}

\begin{figure}[t]
   \plotone{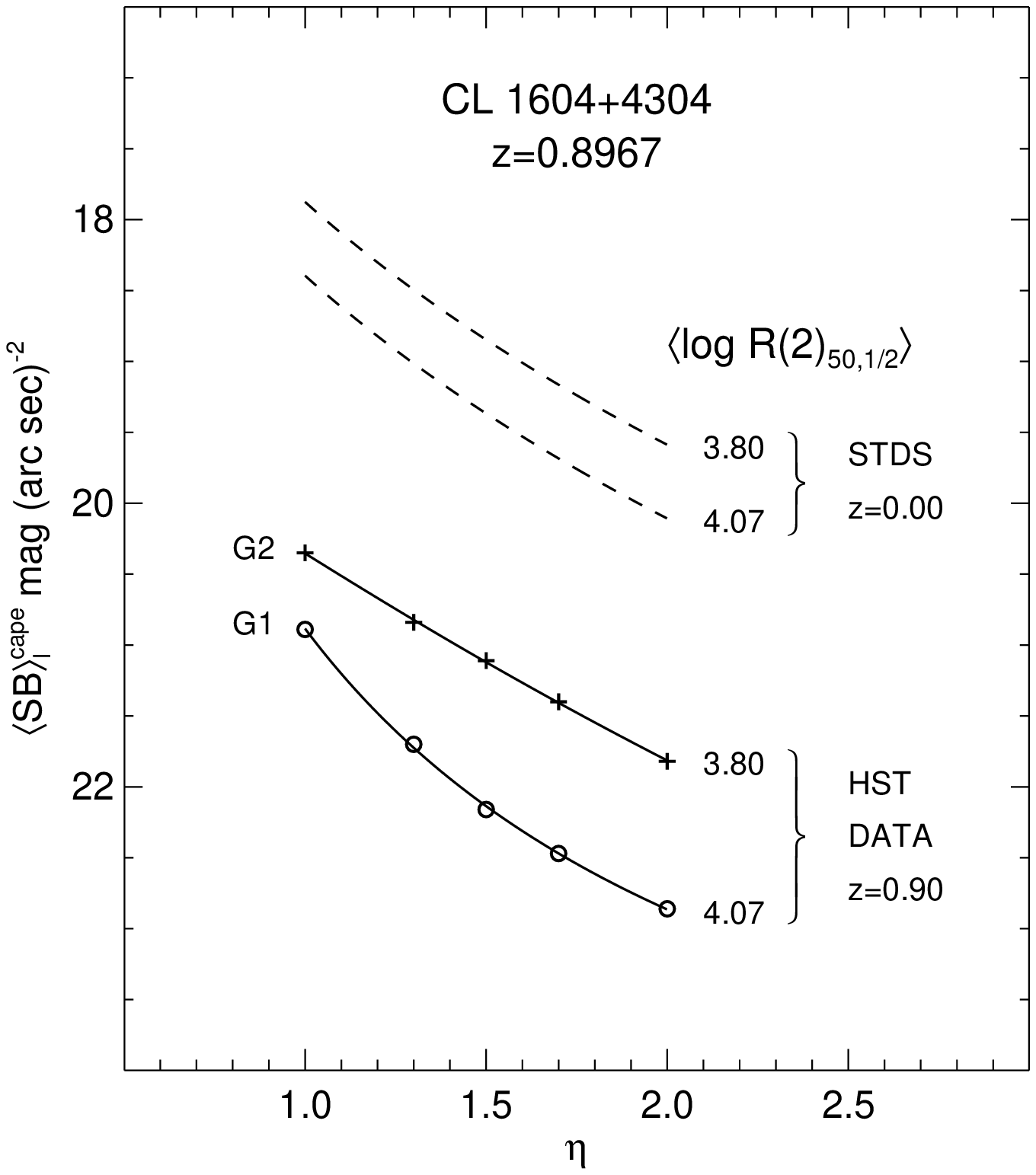}
   \caption{Same as Figure~\ref{fig:08} but for Cl 1604+4304. 
      Data are from Tables~\ref{tab:11} \& \ref{tab:12}.}
\label{fig:10}
\end{figure}

\begin{figure}[t]
   \plotone{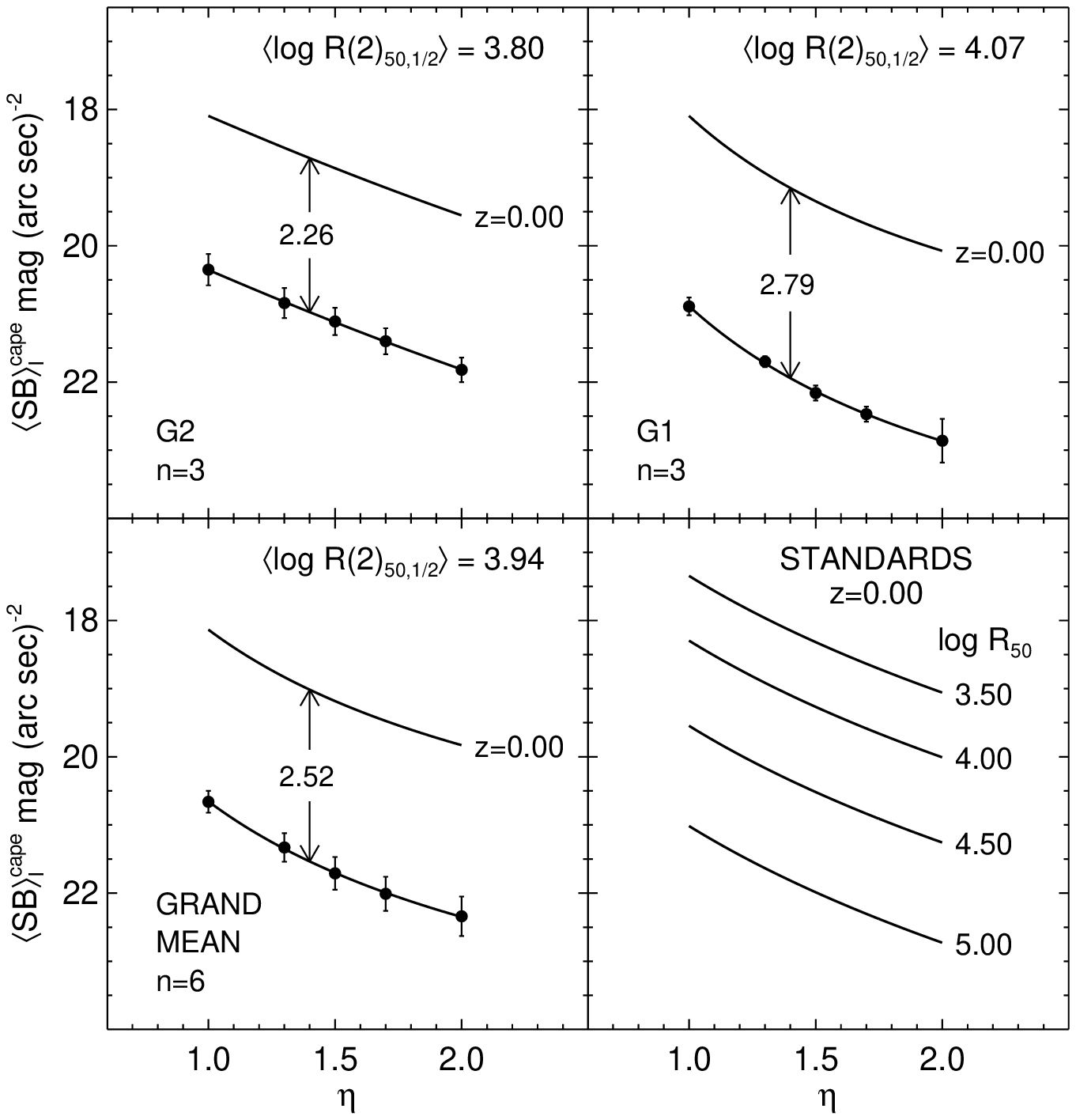}
   \caption{Same as Figure~\ref{fig:09} but for Cl 1604+4304.
      Data are from Tables~\ref{tab:11} \& \ref{tab:12}.}
\label{fig:11}
\end{figure}

\begin{figure}[t]
   \plotone{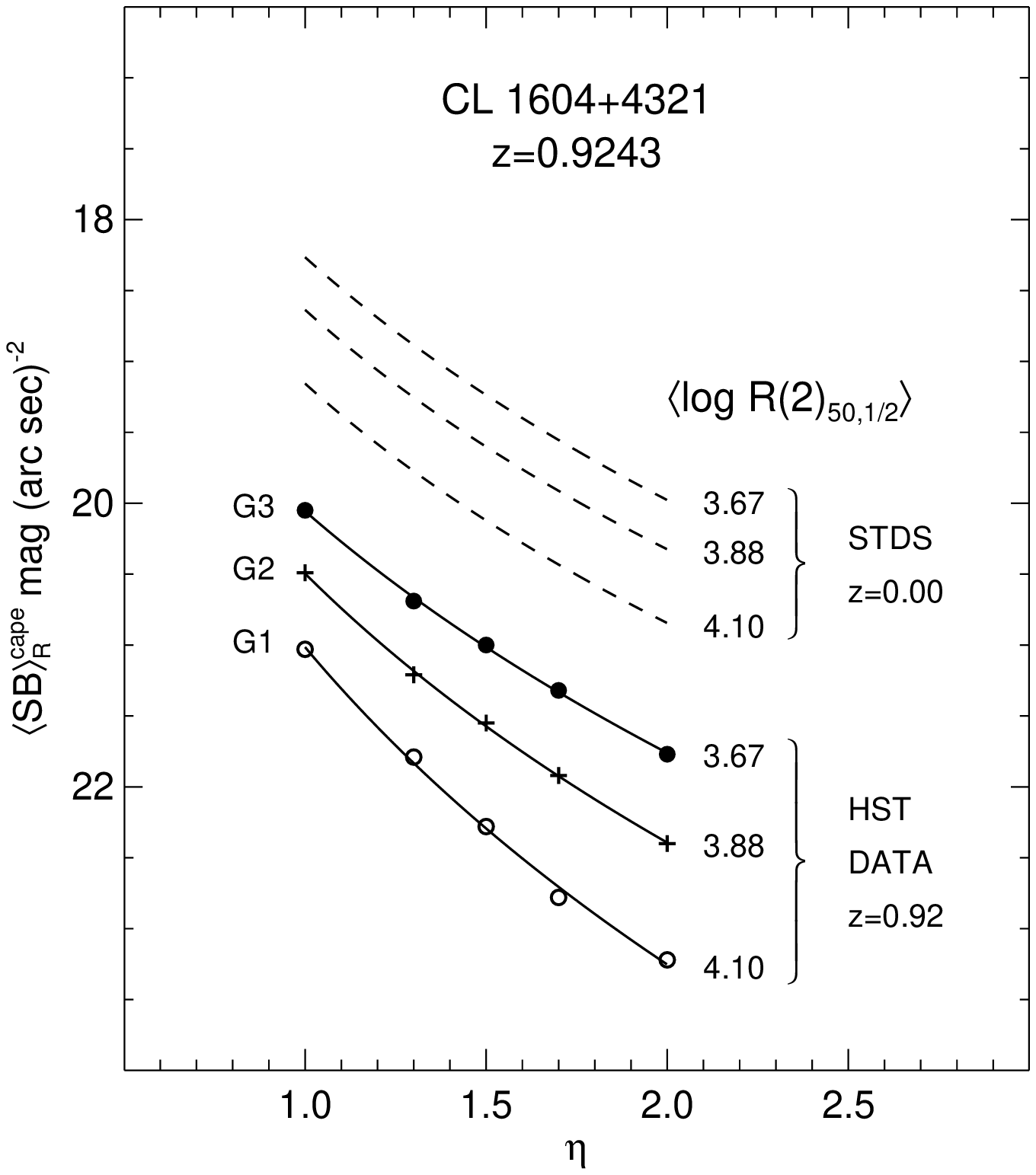}
   \caption{Same as Figure~\ref{fig:08} but for Cl 1604+4321.
      Data are from Tables~\ref{tab:13} \& \ref{tab:14}.}
\label{fig:12}
\end{figure}

\begin{figure}[t]
   \plotone{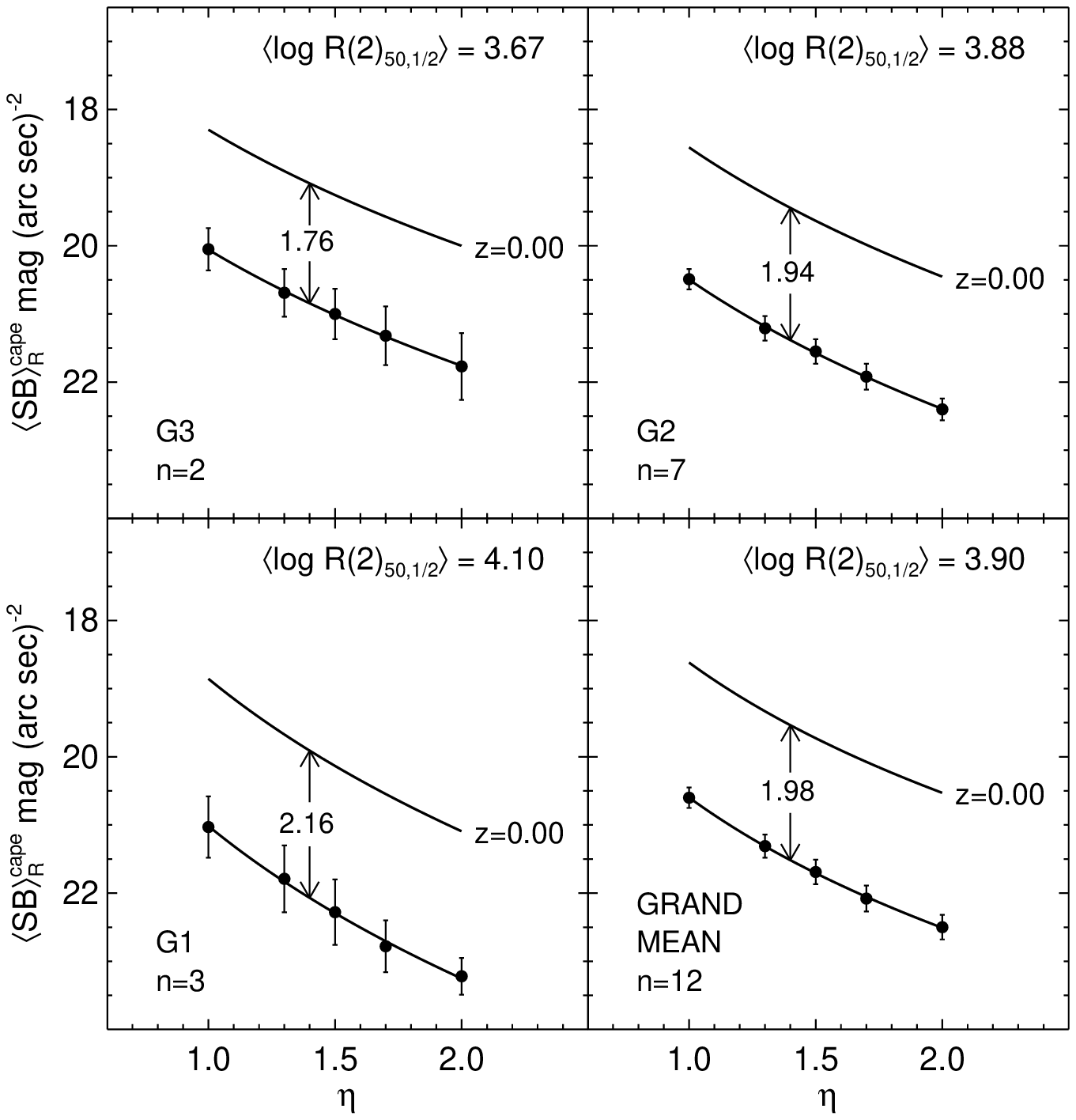}
   \caption{Same as Figure~\ref{fig:09} but for Cl 1604+4321.
      Data are from Tables~\ref{tab:13} \& \ref{tab:14}.}
\label{fig:13}
\end{figure}

\begin{figure}[t]
   \plotone{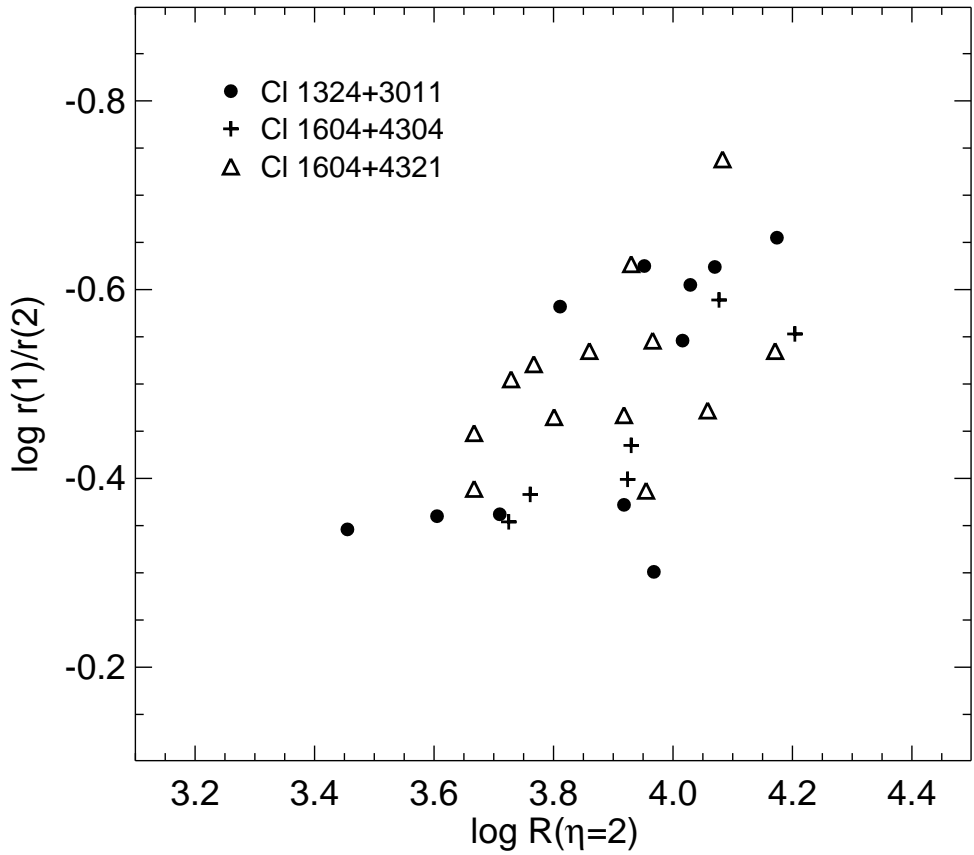}
   \caption{Variation of $\log R(1)/R(2)$ with $\log R(2)$ for galaxies
      in the three remote \textsl{HST} clusters. 
      Dots are for Cl 1324+3011, 
      roman crosses are for Cl 1604+4304,
      open triangles are for Cl 1604+4321. 
      Data are calculated from Tables~2--4 of \citeauthor{LS:01c}.}
\label{fig:14}
\end{figure}

     (9). Figures~\ref{fig:08}--\ref{fig:13} show the comparison of
the $\langle$SB$\rangle$ of local E galaxies with the three \textsl{HST}
clusters, broken into radius bins to compensate for the variation of
$\langle$SB$\rangle$ with intrinsic radii. The Tolman signals,
degraded by luminosity evolution, are listed in
Table~\ref{tab:15}. The errors are put at $\pm0.18\mag$ in each radius
bin by the accounting in \S~\ref{sec:06:a}. 

     (10). Equations~(\ref{eq:10}) and (\ref{eq:11}) divide the
observed data between a Tolman $(1+z)^{4}$ signal and that portion
due to luminosity evolution.

\begin{figure}[t]
   \plotone{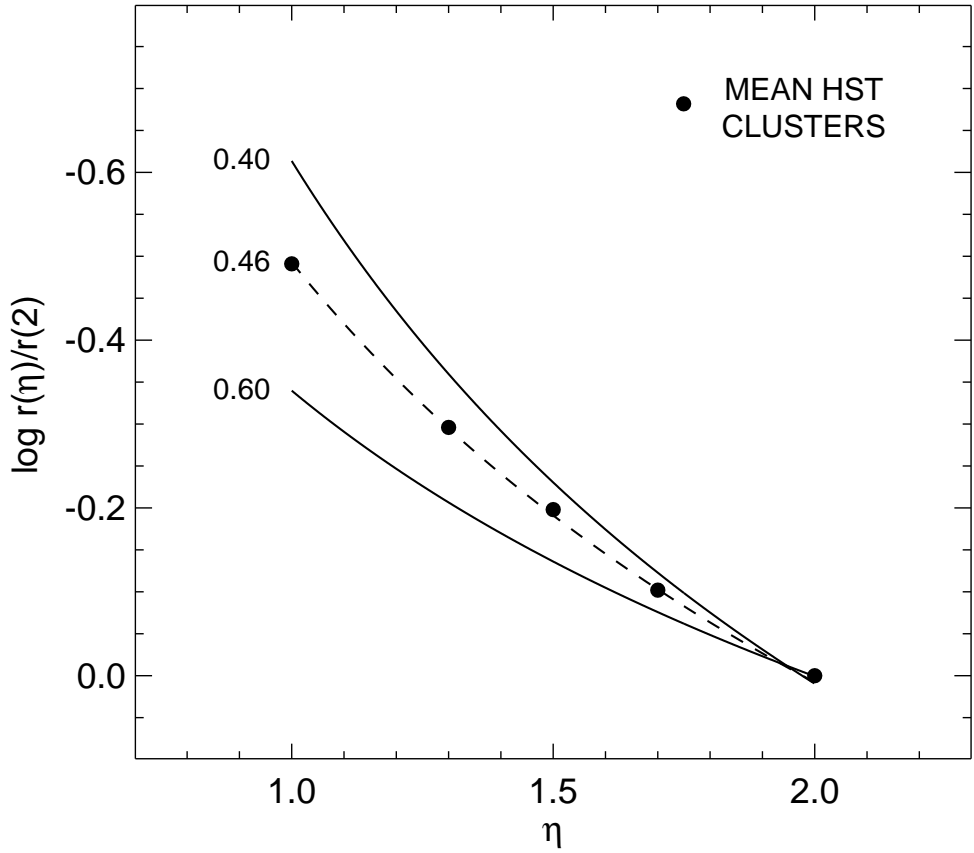}
   \caption{The run of the observed $\eta$ values with $\log r(\eta)/r(2)$ 
       averaged over all intrinsic radii for the galaxies in the
       \textsl{HST} clusters. The S{\'e}rsic functions for $n$ between 0.4
       and 0.6, from Table~\ref{tab:04}, are shown for comparison. 
       From this, we deduce that the mean S{\'e}rsic exponent, averaged
       all radii for the \textsl{HST} cluster galaxies, is $n=0.46$.}
\label{fig:15}
\end{figure}

     (11). That a S{\'e}rsic profile with $n=0.46$ is appropriate for the
local standards and for the three \textsl{HST} cluster galaxies at these
absolute magnitudes, is shown in Figures~\ref{fig:06}, \ref{fig:07} \&
\ref{fig:15}, using Tables~\ref{tab:04} \& \ref{tab:16}. 

     The two conclusions are that the universe expands, and that there
is luminosity evolution in the look-back time. Although  Q.E.D., it
has not been quite so easily done as the way we tried to sell it for
the \textsl{LST} in 1974.

% ***********************************************
%  Acknowledgments
% ***********************************************
\acknowledgments
I am grateful to G.~A. Tammann for reading an early draft of the paper
and for making comments that have clarified a number of the arguments. 
Bernd Reindl's skill is greatly appreciated in preparing the diagrams
in digital form, and in preparing the text in the proper format with
dispatch. 
John Grula, editorial chief for the Observatories, formed again the
liaison with the press, for which I am grateful. 
I thank the Carnegie Institution for its support with post retirement
facilities and publication charges.

% ******************************************************************
% Bibliography
% ******************************************************************

% ******************************************************************

% ******************************************************************
% ***********              Tables                        ***********
% ******************************************************************
\clearpage
\setlength\textwidth{7.01in}  % for Table 1 => no blank page behind

% ******************************************************************
%  Table 1: Growth-Curve Magnitudes and Petrosian eta Radii as 
%           Function of log r/r_e For Various Hubble Types and 
%           for the de Vaucouleurs r^{1/4} law. 
% ******************************************************************
\begin{deluxetable}{rc@{}cc@{}cc@{}cc@{}cc@{}cc@{}cc@{}cc@{}c}
\rotate
\tablewidth{0pt}
\tabletypesize{\footnotesize}
\tablecaption{Growth-Curve Magnitudes and Petrosian $\eta$ Radii as
        Function of $\log r/r_{e}$ For Various Hubble Types and for
        the de~Vaucouleurs $r^{1/4}$ law.\label{tab:01}}  
% ***********************************************
\tablehead{
% ***********************************************
 & & &
   \multicolumn{2}{c}{S{\'e}rsic} & & &
   \multicolumn{2}{c}{S{\'e}rsic} & & & & & & & 
\\
\cline{4-5}
\cline{8-9}
                                         & 
 \multicolumn{2}{c}{$T=-5:\;$E}          &
 \multicolumn{2}{c}{$n=0.25$}            &
 \multicolumn{2}{c}{$T=1:\;$Sa}          &
 \multicolumn{2}{c}{$n=0.46$}            &
 \multicolumn{2}{c}{$T=3:\;$Sb}          &
 \multicolumn{2}{c}{$T=5:\;$Sc}          & 
 \multicolumn{2}{c}{$T=7:\;$Sd}          &
 \multicolumn{2}{c}{$T=9:\;$Sm} 
\\
 \colhead{$\log r/r_{e}$}                &
 \colhead{$\Delta m$} & \colhead{$\eta$} &
 \colhead{$\Delta m$} & \colhead{$\eta$} &
 \colhead{$\Delta m$} & \colhead{$\eta$} &
 \colhead{$\Delta m$} & \colhead{$\eta$} &
 \colhead{$\Delta m$} & \colhead{$\eta$} &
 \colhead{$\Delta m$} & \colhead{$\eta$} &
 \colhead{$\Delta m$} & \colhead{$\eta$} &
 \colhead{$\Delta m$} & \colhead{$\eta$}
\\
 \colhead{(1)}                           &  
 \colhead{(2)}  & \colhead{(3)}          & 
 \colhead{(4)}  & \colhead{(5)}          &
 \colhead{(6)}  & \colhead{(7)}          &
 \colhead{(8)}  & \colhead{(9)}          &
 \colhead{(10)} & \colhead{(11)}         &
 \colhead{(12)} & \colhead{(13)}         &
 \colhead{(14)} & \colhead{(15)}         &
 \colhead{(16)} & \colhead{(17)}            
}     
% ***********************************************
\startdata
% ***********************************************
 $-$1.0 &   2.97 &\nodata &        &        &   3.22 &\nodata &        &        &   3.41 &\nodata &   3.65 &\nodata &   3.99 &\nodata &   4.43 &\nodata\\
 $-$0.9 &   2.69 & 0.65   &   2.70 & 0.61   &   2.95 & 0.67   &   3.45 & 0.36   &   3.13 & 0.61   &   3.33 & 0.54   &   3.63 & 0.39   &   4.01 & 0.22  \\
 $-$0.8 &   2.42 & 0.69   &   2.36 & 0.77   &   2.68 & 0.61   &   3.00 & 0.40   &   2.84 & 0.59   &   3.04 & 0.59   &   3.29 & 0.48   &   3.61 & 0.27  \\
 $-$0.7 &   2.16 & 0.73   &   2.04 & 0.83   &   2.38 & 0.61   &   2.62 & 0.44   &   2.55 & 0.67   &   2.77 & 0.67   &   2.99 & 0.57   &   3.23 & 0.34  \\
 $-$0.6 &   1.91 & 0.80   &   1.80 & 0.89   &   2.11 & 0.73   &   2.30 & 0.49   &   2.30 & 0.71   &   2.50 & 0.59   &   2.70 & 0.47   &   2.88 & 0.28  \\
 $-$0.5 &   1.68 & 0.89   &   1.61 & 0.97   &   1.87 & 0.75   &   2.00 & 0.55   &   2.03 & 0.65   &   2.19 & 0.56   &   2.34 & 0.40   &   2.46 & 0.27  \\
 $-$0.4 &   1.47 & 0.94   &   1.41 & 1.03   &   1.61 & 0.80   &   1.69 & 0.62   &   1.75 & 0.69   &   1.90 & 0.57   &   2.01 & 0.47   &   2.10 & 0.34  \\
 $-$0.3 &   1.26 & 1.02   &   1.23 & 1.10   &   1.39 & 0.92   &   1.42 & 0.72   &   1.50 & 0.73   &   1.60 & 0.56   &   1.68 & 0.45   &   1.73 & 0.34  \\
 $-$0.2 &   1.08 & 1.14   &   1.06 & 1.19   &   1.18 & 0.92   &   1.21 & 0.82   &   1.24 & 0.73   &   1.30 & 0.59   &   1.35 & 0.48   &   1.37 & 0.42  \\
 $-$0.1 &   0.91 & 1.20   &   0.90 & 1.28   &   0.96 & 0.92   &   0.92 & 0.94   &   0.99 & 0.78   &   1.02 & 0.65   &   1.04 & 0.65   &   1.05 & 0.52  \\
    0.0 &   0.75 & 1.31   &   0.75 & 1.39   &   0.75 & 0.89   &   0.74 & 1.08   &   0.75 & 0.89   &   0.75 & 0.73   &   0.75 & 0.69   &   0.75 & 0.65  \\
    0.1 &   0.61 & 1.46   &   0.64 & 1.50   &   0.57 & 1.14   &   0.60 & 1.23   &   0.55 & 1.05   &   0.53 & 1.00   &   0.51 & 0.94   &   0.50 & 0.89  \\
    0.2 &   0.49 & 1.64   &   0.50 & 1.63   &   0.40 & 1.38   &   0.45 & 1.41   &   0.37 & 1.31   &   0.35 & 1.27   &   0.33 & 1.31   &   0.31 & 1.27  \\
    0.3 &   0.39 & 1.80   &   0.41 & 1.80   &   0.29 & 1.69   &   0.31 & 1.61   &   0.25 & 1.64   &   0.22 & 1.60   &   0.21 & 1.60   &   0.19 & 1.64  \\
    0.4 &   0.30 & 1.99   &   0.32 & 1.97   &   0.19 & 1.92   &   0.22 & 1.87   &   0.15 & 1.99   &   0.12 & 2.06   &   0.10 & 2.06   &   0.09 & 2.14  \\
    0.5 &   0.23 & 2.12   &   0.25 & 2.14   &   0.12 & 2.22   &   0.15 & 2.19   &   0.09 & 2.40   &   0.07 & 2.61   &   0.06 & 2.89   &   0.05 & 2.89  \\
    0.6 &   0.17 & 2.50   &   0.18 & 2.38   &   0.06 & 2.89   &   0.09 & 2.58   &   0.04 & 3.06   &   0.03 & 3.25   &   0.03 & 3.25   &   0.02 & 3.50  \\
    0.7 &   0.13 & 2.24   &   0.14 & 2.60   &   0.05 & 3.81   &   0.03 & 3.00   &   0.03 & 4.25   &   0.02 & 4.25   &   0.01 & 3.81   &   0.01 & 4.25  \\
    0.8 &   0.09 & 2.89   &   0.10 & 2.90   &   0.03 & 3.50   &   0.01 & 3.52   &   0.02 &\nodata &   0.01 &\nodata &   0.01 & 4.25   &        &       \\
    0.9 &   0.06 & 3.25   &   0.07 & 3.27   &   0.01 & 4.25   &   0.00 & 4.20   &   0.01 &\nodata &        &        &        &        &        &       \\
    1.0 &   0.04 & 3.81   &   0.05 & 3.70   &   0.00 &\nodata &        &        &        &        &        &        &        &        &        &       \\
% ***********************************************
\enddata
% ***********************************************
\end{deluxetable}
% ***********************************************

\clearpage
\setlength\textwidth{6.5in}   % orig

% ******************************************************************
%  Table 2: Calculated b_n Values for the Sersic Family of
%           Luminosity Profiles.  
% ******************************************************************
\begin{deluxetable}{cccc}
\tablewidth{0pt}
\tabletypesize{\footnotesize}
\tablecaption{Calculated $b_{n}$ Values for the S{\'e}rsic 
              Family of Luminosity Profiles.\label{tab:02}} 
% ***********************************************
\tablehead{
% ***********************************************
 \colhead{S{\'e}rsic $n$} &
 \colhead{$b_{n}$}        & 
 \colhead{S{\'e}rsic $n$} &
 \colhead{$b_{n}$}          
\\
 \colhead{(1)}            &  
 \colhead{(2)}            &
 \colhead{(3)}            & 
 \colhead{(4)}
}     
% ***********************************************
\startdata
% ***********************************************
   0.15 & 5.62 & 0.46 & 1.72 \\
   0.20 & 4.17 & 0.60 & 1.28 \\
   0.25 & 3.33 & 1.00 & 0.73 \\
   0.40 & 2.00 & 1.50 & 0.44 \\
% ***********************************************
\enddata
% ***********************************************
\end{deluxetable}
% ***********************************************

\clearpage

% ******************************************************************
%  Table 3: Growth Curves and eta Values for the Sersic Function at
%           the Listed log r/r_e Ratios.  
% ******************************************************************
\begin{deluxetable}{rc@{}cc@{}cc@{}cc@{}cc@{}cc@{}cc@{}cc@{}c}
\rotate
\tablewidth{0pt}
\tabletypesize{\footnotesize}
\tablecaption{Growth Curves and $\eta$ Values for the S{\'e}rsic 
            Function at the Listed $\log r/r_{e}$ Ratios.\label{tab:03}}
% ***********************************************
\tablehead{
% ***********************************************
                              &
 \multicolumn{2}{c}{$n=0.15$} &
 \multicolumn{2}{c}{$n=0.20$} &
 \multicolumn{2}{c}{$n=0.25$} &
 \multicolumn{2}{c}{$n=0.40$} &
 \multicolumn{2}{c}{$n=0.46$} &
 \multicolumn{2}{c}{$n=0.60$} &
 \multicolumn{2}{c}{$n=1.00$} &
 \multicolumn{2}{c}{$n=1.50$}       
\\
 \colhead{$\log r/r_{e}$}     &
 \colhead{$\Delta m$}         &
 \colhead{$\eta$}             & 
 \colhead{$\Delta m$}         &
 \colhead{$\eta$}             & 
 \colhead{$\Delta m$}         &
 \colhead{$\eta$}             & 
 \colhead{$\Delta m$}         &
 \colhead{$\eta$}             &   
 \colhead{$\Delta m$}         &
 \colhead{$\eta$}             & 
 \colhead{$\Delta m$}         &
 \colhead{$\eta$}             & 
 \colhead{$\Delta m$}         &
 \colhead{$\eta$}             & 
 \colhead{$\Delta m$}         &
 \colhead{$\eta$}          
\\
 \colhead{(1)}                &  
 \colhead{(2)}                &
 \colhead{(3)}                & 
 \colhead{(4)}                &
 \colhead{(5)}                &
 \colhead{(6)}                & 
 \colhead{(7)}                &
 \colhead{(8)}                &
 \colhead{(9)}                & 
 \colhead{(10)}               &
 \colhead{(11)}               & 
 \colhead{(12)}               &
 \colhead{(13)}               & 
 \colhead{(14)}               &
 \colhead{(15)}               & 
 \colhead{(16)}               &
 \colhead{(17)}     
}
% ***********************************************
\startdata
% ***********************************************
 $-$0.4 &   1.26 & 1.26   &   1.34 & 1.13   &   1.41 & 1.03   &    1.60 & 0.74   &        &        &   1.79 &\nodata &   2.05 &\nodata &   2.30 &\nodata \\
 $-$0.3 &   1.14 & 1.35   &   1.19 & 1.20   &   1.23 & 1.10   &    1.32 & 0.80   &        &        &   1.55 &\nodata &   1.70 &\nodata &   1.81 &\nodata \\
 $-$0.2 &   1.00 & 1.41   &   1.02 & 1.29   &   1.06 & 1.19   &    1.10 & 0.88   &   1.18 & 0.82   &   1.28 & 0.66   &   1.40 & 0.40   &   1.42 & 0.21   \\
 $-$0.1 &   0.86 & 1.51   &   0.89 & 1.38   &   0.90 & 1.28   &    0.94 & 1.00   &   0.92 & 0.94   &   1.00 & 0.76   &   1.05 & 0.54   &   1.08 & 0.32   \\
    0.0 &   0.75 & 1.60   &   0.75 & 1.47   &   0.75 & 1.39   &    0.75 & 1.14   &   0.74 & 1.08   &   0.75 & 0.90   &   0.75 & 0.72   &   0.75 & 0.50   \\
    0.1 &   0.64 & 1.70   &   0.63 & 1.60   &   0.64 & 1.50   &    0.61 & 1.26   &   0.59 & 1.23   &   0.55 & 1.10   &   0.50 & 0.94   &   0.44 & 0.73   \\
    0.2 &   0.56 & 1.80   &   0.52 & 1.71   &   0.50 & 1.63   &    0.48 & 1.44   &   0.45 & 1.41   &   0.38 & 1.30   &   0.32 & 1.20   &   0.25 & 1.06   \\
    0.3 &   0.48 & 1.91   &   0.43 & 1.88   &   0.41 & 1.80   &    0.33 & 1.63   &   0.31 & 1.61   &   0.26 & 1.56   &   0.20 & 1.52   &   0.11 & 1.53   \\
    0.4 &   0.40 & 2.08   &   0.36 & 2.01   &   0.32 & 1.97   &    0.22 & 1.86   &   0.22 & 1.87   &   0.15 & 1.90   &   0.11 & 2.10   &   0.04 & 2.36   \\
    0.5 &   0.33 & 2.19   &   0.29 & 2.19   &   0.25 & 2.14   &    0.16 & 2.14   &   0.15 & 2.19   &   0.08 & 2.31   &   0.04 & 2.80   &   0.01 & 4.10   \\
    0.6 &   0.27 & 2.32   &   0.21 & 2.37   &   0.18 & 2.38   &    0.10 & 2.43   &   0.09 & 2.58   &   0.03 & 2.80   &   0.00 & 3.91   &        &     \\
    0.7 &   0.21 & 2.51   &   0.16 & 2.53   &   0.14 & 2.60   &    0.06 & 2.80   &   0.03 & 3.00   &   0.00 & 3.38   &        &        &        &     \\
    0.8 &   0.15 & 2.67   &   0.11 & 2.77   &   0.10 & 2.90   &    0.04 & 3.24   &   0.01 & 3.52   &        &        &        &        &        &     \\
    0.9 &   0.11 & 2.88   &   0.09 & 2.99   &   0.07 & 3.27   &    0.02 & 3.75   &   0.00 & 4.20   &        &        &        &        &        &     \\
    1.0 &   0.07 & 3.10   &   0.04 & 3.23   &   0.05 & 3.70   &    0.00 &\nodata &        &        &        &        &        &        &        &     \\
    1.1 &   0.05 & 3.32   &   0.02 & 3.53   &   0.03 &\nodata &    0.00 &\nodata &        &        &        &        &        &        &        &     \\
    1.2 &   0.02 & 3.50   &   0.01 & 3.88   &        &        &         &        &        &        &        &        &        &        &        &     \\
% ***********************************************
\enddata
% ***********************************************
\end{deluxetable}
% ***********************************************
                                                                                                                                               
\clearpage
                                                                                                                                                        
% ******************************************************************
%  Table 4: log r/r(eta=2) at Various Eta Values for the Sersic 
%           Family from Table 3.
% ******************************************************************
\begin{deluxetable}{lrrrrrrrrr}
\tablewidth{0pt}
\tabletypesize{\footnotesize}
\tablecaption{$\log r/r(\eta=2)$ at Various $\eta$ Values for the S{\'e}rsic 
                   Family from Table~\ref{tab:03}.\label{tab:04}}
% ***********************************************
\tablehead{
% ***********************************************
 \colhead{$\eta/n$} &  
 \colhead{0.15}     &
 \colhead{0.20}     & 
 \colhead{0.25}     &
 \colhead{$T=-5$}   &
 \colhead{0.40}     & 
 \colhead{0.46}     &
 \colhead{0.60}     &
 \colhead{1.00}     & 
 \colhead{1.50}
\\
 \colhead{(1)}      &  
 \colhead{(2)}      &
 \colhead{(3)}      & 
 \colhead{(4)}      &
 \colhead{(5)}      &
 \colhead{(6)}      & 
 \colhead{(7)}      &
 \colhead{(8)}      &
 \colhead{(9)}      & 
 \colhead{(10)}
}
% ***********************************************
\startdata
% ***********************************************
  1.0 &\nodata &($-$1.0) & $-$0.82 & $-$0.72 & $-$0.65 & $-$0.50 & $-$0.34 & $-$0.27 & $-$0.15 \\
  1.3 &$-$0.78 & $-$0.58 & $-$0.51 & $-$0.42 & $-$0.31 & $-$0.29 & $-$0.21 & $-$0.15 & $-$0.10 \\
  1.5 &$-$0.51 & $-$0.38 & $-$0.32 & $-$0.28 & $-$0.21 & $-$0.18 & $-$0.14 & $-$0.09 & $-$0.06 \\
  1.7 &$-$0.29 & $-$0.21 & $-$0.20 & $-$0.16 & $-$0.11 & $-$0.10 & $-$0.06 & $-$0.06 & $-$0.03 \\
  2.0 &  0.00  &    0.00 &    0.00 &    0.00 &    0.00 &    0.00 &    0.00 &    0.00 &    0.00 \\
  2.5 &  0.33  &    0.28 &    0.22 &    0.21 &    0.17 &    0.15 &    0.09 &    0.07 &    0.05 \\
  3.0 &  0.59  &    0.50 &    0.41 &    0.40 &    0.28 &    0.26 &    0.18 &    0.13 &    0.09 \\
  3.5 &  0.81  &    0.70 &    0.56 &    0.54 &    0.40 &    0.36 &    0.26 &    0.19 &    0.13 \\
  4.0 &\nodata &    0.83 &    0.69 &    0.62 &    0.50 &    0.44 &    0.30 &    0.22 &    0.17 \\
% ***********************************************
\enddata
% ***********************************************
\end{deluxetable}
% ***********************************************

\clearpage

% ******************************************************************
%  Table 5: log r/r(eta=2) as Function of eta and the alpha-to-beta 
%           Ratio for Four Oemler Profiles and the Observed T = -5 and 
%           a Sersic Profile With n=0.46.
% ******************************************************************
\begin{deluxetable}{rrrrrrr}
\tablewidth{0pt}
\tabletypesize{\footnotesize}
\tablecaption{$\log r/r(\eta=2)$ as Function of $\eta$ and the $\alpha$-to-$\beta$ 
              Ratio for Four Oemler Profiles and the Observed $T=-5$ and a 
              S{\'e}rsic Profile With $n=0.46$.\label{tab:05}}
% ***********************************************
\tablehead{
% ***********************************************
                                               &
 \multicolumn{4}{c}{$\alpha$-to-$\beta$ Ratio} &
 \multicolumn{1}{c}{E} &
 \multicolumn{1}{c}{S{\'e}rsic}
\\
\cline{2-5}
 \colhead{$\eta$} &  
 \colhead{10}     &
 \colhead{30}     &
 \colhead{60}     &
 \colhead{100}    &
 \colhead{$T=-5$} &
 \colhead{0.46} 
\\
 \colhead{(1)}    &  
 \colhead{(2)}    &
 \colhead{(3)}    & 
 \colhead{(4)}    &
 \colhead{(5)}    &
 \colhead{(6)}    & 
 \colhead{(7)}
}
% ***********************************************
\startdata
% ***********************************************
 1.0 & $-$0.44 & $-$0.70 & $-$0.88 & $-$0.91 & $-$0.72 & $-$0.50 \\
 1.3 & $-$0.25 & $-$0.42 & $-$0.58 & $-$0.67 & $-$0.42 & $-$0.29 \\
 1.5 & $-$0.18 & $-$0.26 & $-$0.38 & $-$0.46 & $-$0.28 & $-$0.18 \\
 1.7 & $-$0.10 & $-$0.14 & $-$0.21 & $-$0.25 & $-$0.16 & $-$0.10 \\
 2.0 &    0.00 &    0.00 &    0.00 &    0.00 &    0.00 &    0.00 \\
 2.5 &    0.10 &    0.15 &    0.20 &    0.26 &    0.21 &    0.15 \\
 3.0 &    0.18 &    0.24 &    0.31 &    0.40 &    0.40 &    0.26 \\
 3.5 &    0.22 &    0.31 &    0.39 &    0.48 &    0.54 &    0.36 \\
% ***********************************************
\enddata
% ***********************************************
\end{deluxetable}
% ******************************************************************

% ******************************************************************
%  Table 6: Surface Brightness Normalized to eta=2 for Eight
%           Sersic Profiles, Averaged Over the Area Interior to the 
%           Listed Eta Radii. The Unit is Magnitudes (Relative to
%           eta=2 Values) Per Unit Area. 
% ******************************************************************
\begin{deluxetable}{lrrrrrrrrr}
\tablewidth{0pt}
\tabletypesize{\footnotesize}
\tablecaption{Surface Brightness Normalized to $\eta=2$ for Eight
   S{\'e}rsic Profiles, Averaged Over the Area Interior to the Listed $\eta$
   Radii. The Unit is Magnitudes (Relative to $\eta=2$ Values) Per
   Unit Area.\label{tab:06}}
% ***********************************************
\tablehead{
% ***********************************************
 \colhead{$\eta/n$} &  
 \colhead{0.15}     &
 \colhead{0.20}     & 
 \colhead{0.25}     &
 \colhead{$T=-5$\tablenotemark{a}}   &
 \colhead{0.40}     & 
 \colhead{0.46}     &
 \colhead{0.60}     &
 \colhead{1.00}     & 
 \colhead{1.50}
\\
 \colhead{(1)}      &  
 \colhead{(2)}      &
 \colhead{(3)}      & 
 \colhead{(4)}      &
 \colhead{(5)}      &
 \colhead{(6)}      & 
 \colhead{(7)}      &
 \colhead{(8)}      &
 \colhead{(9)}      & 
 \colhead{(10)}
}
% ***********************************************
\startdata
% ***********************************************
  1.0 & \nodata & $-$3.30 & $-$2.82 & $-$2.78 & $-$2.02 & $-$1.70 & $-$1.38 & $-$0.93 & $-$0.52 \\
  1.3 & \nodata & $-$2.15 & $-$1.82 & $-$1.78 & $-$1.18 & $-$1.10 & $-$0.83 & $-$0.58 & $-$0.35 \\
  1.5 & $-$1.80 & $-$1.50 & $-$1.20 & $-$1.25 & $-$0.78 & $-$0.73 & $-$0.60 & $-$0.38 & $-$0.26 \\
  1.7 & $-$1.07 & $-$0.85 & $-$0.68 & $-$0.73 & $-$0.48 & $-$0.40 & $-$0.38 & $-$0.20 & $-$0.14 \\
  2.0 &    0.00 &    0.00 &    0.00 &    0.00 &    0.00 &    0.00 &    0.00 &    0.00 &    0.00 \\
  2.5 &    1.52 &    1.37 &    1.10 &    0.97 &    0.68 &    0.64 &    0.48 &    0.32 &    0.25 \\
  3.0 &    2.66 &    2.45 &    1.90 &    1.58 &    1.30 &    1.25 &    1.13 &    0.59 &    0.46 \\  
  3.5 &    3.62 &    3.30 &    2.60 &    2.20 &    1.85 &    1.72 &    1.54 &    0.80 &    0.66 \\
  4.0 & \nodata &    3.90 & \nodata &    2.72 &    2.30 &    2.21 &    2.05 &    1.00 &    0.84 \\
% ***********************************************
\enddata
% ***********************************************
\tablenotetext{a}{Calculated from the observed $T=-5$ growth curve.}
% ***********************************************
\end{deluxetable}
% ***********************************************

% ******************************************************************
%  Table 7: Nonliner Corrections to Equations (6)-(9). 
% ******************************************************************
\begin{deluxetable}{ccccc}
\tablewidth{0pt}
\tabletypesize{\footnotesize}
\tablecaption{Nonlinear Corrections to Equations~(\ref{eq:06})--(\ref{eq:09}).\tablenotemark{a}\label{tab:07}} 
% ***********************************************
\tablehead{
% ***********************************************
 \colhead{$\log R$}                   &
 \colhead{$\Delta\langle$SB$\rangle$} & &
 \colhead{$\log R$}                   &
 \colhead{$\Delta\langle$SB$\rangle$}
\\
                                      &
 \colhead{(mag)}                      & &
                                      &
 \colhead{(mag)}
\\
 \colhead{(1)}                        &  
 \colhead{(2)}                        & &
 \colhead{(3)}                        & 
 \colhead{(4)}
}     
% ***********************************************
\startdata
% ***********************************************
   4.4 & 0.00   & &  3.8 & 0.42 \\ 
   4.3 & 0.02   & &  3.7 & 0.50 \\
   4.2 & 0.06   & &  3.6 & 0.60 \\
   4.1 & 0.10   & &  3.5 & 0.67 \\
   4.0 & 0.23   & &  3.4 & 0.75 \\
   3.9 & 0.36   & &  3.3 & 0.86 \\
% ***********************************************
\enddata
% ***********************************************
\tablenotetext{a}{The $R$ radii are based on distance moduli of 31.7 for Virgo, 
                   31.9 for Fornax, and 35.5 for Coma.}
% ***********************************************
\end{deluxetable}
% ***********************************************

% ******************************************************************
%  Table 8: Ridge-Line Variation of <SB> with the log R Intrinsic Size
%           at eta=2 at Zero Redshift for E Galaxies in the
%           V(RI)_Cape Pass Bands
% ******************************************************************
\begin{deluxetable}{cccc}
\tablewidth{0pt}
\tabletypesize{\footnotesize}
\tablecaption{Ridge-Line Variation of $\langle$SB$\rangle$ with the $\log R$ Intrinsic Size 
              at $\eta=2$ at Zero Redshift for E Galaxies in the $V(RI)_{\rm cape}$ 
              Pass Bands.\tablenotemark{a,b}\label{tab:08}} 
% ***********************************************
\tablehead{
% ***********************************************
 \colhead{$\log R(2)$}              &
 \colhead{$\langle$SB$\rangle_{V}$} & 
 \colhead{$\langle$SB$\rangle_{R}$} &
 \colhead{$\langle$SB$\rangle_{I}$}          
\\
 \colhead{(1)}                      &  
 \colhead{(2)}                      &
 \colhead{(3)}                      & 
 \colhead{(4)}
}     
% ***********************************************
\startdata
% ***********************************************
  5.4   &   25.14 & 24.57 & 23.93 \\
  5.2   &   24.56 & 23.99 & 23.35 \\
  5.0   &   23.95 & 23.38 & 22.74 \\
  4.8   &   23.36 & 22.79 & 22.15 \\
  4.6   &   22.76 & 22.19 & 21.55 \\
  4.4   &   22.17 & 21.60 & 20.96 \\
  4.2   &   21.63 & 21.06 & 20.42 \\
  4.0   &   21.21 & 20.64 & 20.00 \\
  3.8   &   20.81 & 20.24 & 19.60 \\
  3.6   &   20.39 & 19.82 & 19.18 \\
  3.4   &   19.95 & 19.38 & 18.74 \\
% ***********************************************
\enddata
% ***********************************************
\tablenotetext{a}{The entries are calculated from equation~(\ref{eq:09}) using the 
                   nonlinear corrections of Table~\ref{tab:07}.}
\tablenotetext{b}{The unit of the $\langle$SB$\rangle$ entries is mag (arc sec)$^{-2}$.}
% ***********************************************
\end{deluxetable}
% ***********************************************

% ******************************************************************
%  Table 9: HST <SB> Data For a Range of eta Values for Cl 1324+3011
%           Binned Into Three Radius Groups.  
% ******************************************************************
\begin{deluxetable}{lrrrrr}
\tablewidth{0pt}
\tabletypesize{\footnotesize}
\tablecaption{\textsl{HST} $\langle$SB$\rangle$ Data For a Range of $\eta$ Values 
              for Cl 1324+3011 Binned Into Three Radius Groups.\label{tab:09}} 
% ***********************************************
\tablehead{
% ***********************************************
 \colhead{Item/$\eta$} &
 \colhead{1.0}         & 
 \colhead{1.3}         &
 \colhead{1.5}         & 
 \colhead{1.7}         &
 \colhead{2.0}          
\\
 \colhead{(1)}         &  
 \colhead{(2)}         &
 \colhead{(3)}         & 
 \colhead{(4)}         &
 \colhead{(5)}         & 
 \colhead{(6)}
}     
% ***********************************************
\startdata
% ***********************************************
\\[-15pt]
\multicolumn{6}{c}{{\em Group 1}  ~~ $n=2$ ~~ $\langle M_{I}\rangle_{\eta=2}=-23.49$}\\[2pt]
\tableline
% ***********************************************
  $\langle$SB$\rangle_{I}$ &     20.17 &     21.28 &     21.88 &     22.42 &     22.86 \\ 
  error                    & $\pm$0.39 & $\pm$0.22 & $\pm$0.14 & $\pm$0.26 & $\pm$0.33 \\   
  $\langle\log R\rangle$   &     3.487 &     3.815 &     3.911 &     4.061 &     4.087 \\
  rms                      &     0.060 &     0.126 &     0.218 &     0.124 &     0.080 \\
% ***********************************************
\tableline\\[-9pt]
\multicolumn{6}{c}{{\em Group 2}  ~~ $n=6$ ~~ $\langle M_{I}\rangle_{\eta=2}=-22.98$}\\[2pt]
\tableline
% ***********************************************
  $\langle$SB$\rangle_{I}$ &     19.95 &     20.61 &     20.99 &     21.34 &     21.77 \\
  error                    & $\pm$0.24 & $\pm$0.15 & $\pm$0.10 & $\pm$0.08 & $\pm$0.09 \\
  $\langle\log R\rangle$   &     3.424 &     3.608 &     3.708 &     3.797 &     3.898 \\
  rms                      &     0.159 &     0.125 &     0.111 &     0.112 &     0.117 \\
% ***********************************************
\tableline\\[-9pt]
\multicolumn{6}{c}{{\em Group 3}  ~~ $n=5$ ~~ $\langle M_{I}\rangle_{\eta=2}=-22.01$}\\[2pt]
\tableline
% ***********************************************
  $\langle$SB$\rangle_{I}$ &     19.48 &     20.03 &     20.31 &     20.55 &     20.88 \\
  error                    & $\pm$0.35 & $\pm$0.38 & $\pm$0.37 & $\pm$0.37 & $\pm$0.39 \\ 
  $\langle\log R\rangle$   &     3.177 &     3.326 &     3.396 &     3.454 &     3.530 \\
  rms                      &     0.096 &     0.098 &     0.098 &     0.102 &     0.106 \\
% ***********************************************
\tableline\\[-9pt]
\multicolumn{6}{c}{{\em Grand mean}  ~~ $n=13$ ~~ $\langle M_{I}\rangle_{\eta=2}=-23.03$}\\[2pt]
\tableline
% ***********************************************
  $\langle$SB$\rangle_{I}$ &     19.97 &     20.78 &     21.23 &     21.57 &     21.91 \\
  error                    & $\pm$0.18 & $\pm$0.17 & $\pm$0.18 & $\pm$0.22 & $\pm$0.24 \\
  $\langle\log R\rangle$   &     3.410 &     3.644 &     3.769 &     3.827 &     3.883 \\
  rms                      &     0.155 &     0.206 &     0.260 &     0.238 &     0.216 \\
% ***********************************************
\enddata
% ***********************************************
\end{deluxetable}
% ***********************************************

% ******************************************************************
%  Table 10: Zero Redshift Surface Brightness - eta Relation For a
%            Sersic Profile With n=0.46 at the Mean log R(2) for the
%            Three Radius Groups for HST Cl 1324+3011  
% ******************************************************************
\begin{deluxetable}{ccccc}
\tablewidth{0pt}
\tabletypesize{\footnotesize}
\tablecaption{Zero Redshift Surface Brightness - $\eta$ Relation For a 
          S{\'e}rsic Profile With $n=0.46$ at the Mean $\log R(2)$ for the 
   Three Radius Groups for \textsl{HST} Cl 1324+3011.\tablenotemark{a,b}\label{tab:10}} 
% ***********************************************
\tablehead{
% ***********************************************
 \colhead{$\eta/\langle\log R\rangle$} &
 \colhead{4.087}                       & 
 \colhead{3.898}                       &
 \colhead{3.530}                       &
 \colhead{3.882}
\\
 \colhead{(1)}                         &  
 \colhead{(2)}                         &
 \colhead{(3)}                         & 
 \colhead{(4)}                         & 
 \colhead{(5)}
}     
% ***********************************************
\startdata
% ***********************************************
  1.0 &   18.48 & 18.08 & 17.30 & 18.01 \\
  1.3 &   19.08 & 18.68 & 17.90 & 18.61 \\
  1.5 &   19.45 & 19.05 & 18.27 & 18.98 \\
  1.7 &   19.78 & 19.38 & 18.60 & 19.31 \\
  2.0 &   20.18 & 19.78 & 19.00 & 19.71 \\
      &    G1   &  G2   &  G3   &  mean \\  
% ***********************************************
\enddata
% ***********************************************
\tablenotetext{a}{The listed values are the surface brightness averaged 
     over the area interior to the listed $\eta$ radii.}
\tablenotetext{b}{The units for the surface brightness are mag per (arc 
     sec)$^{2}$ in the Cape $I$ band.}
% ***********************************************
\end{deluxetable}
% ***********************************************

% ******************************************************************
%  Table 11: Same as Table 9 for HST Cl 1604+4304. 
% ******************************************************************
\begin{deluxetable}{lrrrrr}
\tablewidth{0pt}
\tabletypesize{\footnotesize}
\tablecaption{Same as Table~\ref{tab:09} for \textsl{HST} 
              Cl 1604+4304 ($z =0.8967$).\label{tab:11}}  
% ***********************************************
\tablehead{
% ***********************************************
 \colhead{Item/$\eta$} &
 \colhead{1.0}         & 
 \colhead{1.3}         &
 \colhead{1.5}         & 
 \colhead{1.7}         &
 \colhead{2.0}          
\\
 \colhead{(1)}         &  
 \colhead{(2)}         &
 \colhead{(3)}         & 
 \colhead{(4)}         &
 \colhead{(5)}         & 
 \colhead{(6)}
}     
% ***********************************************
\startdata
% ***********************************************
\\[-15pt]
\multicolumn{6}{c}{{\em Group 1}  ~~ $n=3/4$ ~~ $\langle M_{I}\rangle_{\eta=2}=-23.09$}\\[2pt]
\tableline
% ***********************************************
  $\langle$SB$\rangle_{I}$ &     20.89 &     21.70 &     22.16 &     22.47 &     22.86 \\ 
  error                    & $\pm$0.13 & $\pm$0.08 & $\pm$0.11 & $\pm$0.11 & $\pm$0.32 \\   
  $\langle\log R\rangle$   &     3.578 &     3.816 &     3.940 &     4.016 &     4.070 \\
  rms                      &     0.106 &     0.156 &     0.141 &     0.145 &     0.137 \\
% ***********************************************
\tableline\\[-9pt]
\multicolumn{6}{c}{{\em Group 2}  ~~ $n=3$ ~~ $\langle M_{I}\rangle_{\eta=2}=-22.79$}\\[2pt]
\tableline
% ***********************************************
  $\langle$SB$\rangle_{I}$ &     20.35 &     20.84 &     21.11 &     21.40 &     21.82 \\
  error                    & $\pm$0.23 & $\pm$0.22 & $\pm$0.20 & $\pm$0.19 & $\pm$0.18 \\
  $\langle\log R\rangle$   &     3.425 &     3.560 &     3.635 &     3.705 &     3.803 \\
  rms                      &     0.087 &     0.094 &     0.097 &     0.102 &     0.106 \\
% ***********************************************
\tableline\\[-9pt]
\multicolumn{6}{c}{{\em Grand mean}  ~~ $n=6$ ~~ $\langle M_{I}\rangle_{\eta=2}=-22.94$}\\[2pt]
\tableline
% ***********************************************
  $\langle$SB$\rangle_{I}$ &     20.66 &     21.33 &     21.71 &     22.01 &     22.34 \\
  error                    & $\pm$0.16 & $\pm$0.21 & $\pm$0.24 & $\pm$0.25 & $\pm$0.29 \\
  $\langle\log R\rangle$   &     3.512 &     3.706 &     3.809 &     3.883 &     3.937 \\
  rms                      &     0.122 &     0.184 &     0.199 &     0.204 &     0.183 \\
% ***********************************************
\enddata
% ***********************************************
\end{deluxetable}
% ***********************************************

% ******************************************************************
%  Table 12: Standard eta-Surface Brightness Curves at Zero Redshift
%            for the Two Radius Bins for HST Cl 1604+4304 and for
%            Four Arbitrary Bins for Illustration. 
% ******************************************************************
\begin{deluxetable}{cccccccc}
\tablewidth{0pt}
\tabletypesize{\footnotesize}
\tablecaption{Standard $\eta$-Surface Brightness Curves at Zero 
        Redshift for the Two Radius Bins for \textsl{HST} Cl 1604+4304  
          and for Four Arbitrary Bins for Illustration.\label{tab:12}} 
% ***********************************************
\tablehead{
% ***********************************************
 \colhead{$\eta/\langle\log R\rangle$} &
 \colhead{4.070}                       & 
 \colhead{3.803}                       &
 \colhead{3.937}                       &
 \colhead{3.50}                        & 
 \colhead{4.00}                        &
 \colhead{4.50}                        &
 \colhead{5.00}
\\
 \colhead{(1)}                         &  
 \colhead{(2)}                         &
 \colhead{(3)}                         & 
 \colhead{(4)}                         & 
 \colhead{(5)}                         &
 \colhead{(6)}                         & 
 \colhead{(7)}                         & 
 \colhead{(8)}
}     
% ***********************************************
\startdata
% ***********************************************
   1.0   &   18.40 & 17.88 & 18.22 & 17.35 & 18.30 & 19.55 & 21.02 \\
   1.3   &   19.00 & 18.48 & 18.82 & 17.95 & 18.90 & 20.15 & 21.62 \\
   1.5   &   19.37 & 18.85 & 19.19 & 18.32 & 19.27 & 20.52 & 21.99 \\
   1.7   &   19.70 & 19.18 & 19.52 & 18.65 & 19.60 & 20.85 & 22.32 \\
   2.0   &   20.10 & 19.58 & 19.92 & 19.05 & 20.00 & 21.25 & 22.72 \\
         &    G1   &  G2   &  mean &       &       &       &       \\
% ***********************************************
\enddata
% ***********************************************
\end{deluxetable}
% ***********************************************

% ******************************************************************
%  Table 13: Same as Tables 9 & 11 but for HST Cl 1604+4321. 
% ******************************************************************
\begin{deluxetable}{lrrrrr}
\tablewidth{0pt}
\tabletypesize{\footnotesize}
\tablecaption{Same as Tables~\ref{tab:09} and \ref{tab:11} but for 
              \textsl{HST} Cl 1604+4321.\label{tab:13}} 
% ***********************************************
\tablehead{
% ***********************************************
 \colhead{Item/$\eta$} &
 \colhead{1.0}         & 
 \colhead{1.3}         &
 \colhead{1.5}         & 
 \colhead{1.7}         &
 \colhead{2.0}          
\\
 \colhead{(1)}         &  
 \colhead{(2)}         &
 \colhead{(3)}         & 
 \colhead{(4)}         &
 \colhead{(5)}         & 
 \colhead{(6)}
}     
% ***********************************************
\startdata
% ***********************************************
\\[-15pt]
\multicolumn{6}{c}{{\em Group 1}  ~~ $n=3$ ~~ $\langle M_{R}\rangle_{\eta=2}=-22.95$}\\[2pt]
\tableline
% ***********************************************
  $\langle$SB$\rangle_{R}$ &     21.03 &     21.79 &     22.28 &     22.78 &     23.22 \\ 
  error                    & $\pm$0.45 & $\pm$0.49 & $\pm$0.48 & $\pm$0.38 & $\pm$0.27 \\   
  $\langle\log R\rangle$   &     3.611 &     3.741 &     3.872 &     4.001 &     4.104 \\
  rms                      &     0.035 &     0.175 &     0.161 &     0.115 &     0.059 \\
% ***********************************************
\tableline\\[-9pt]
\multicolumn{6}{c}{{\em Group 2}  ~~ $n=7$ ~~ $\langle M_{R}\rangle_{\eta=2}=-22.64$}\\[2pt]
\tableline
% ***********************************************
  $\langle$SB$\rangle_{R}$ &     20.49 &     21.21 &     21.55 &     21.92 &     22.40 \\
  error                    & $\pm$0.15 & $\pm$0.18 & $\pm$0.18 & $\pm$0.19 & $\pm$0.16 \\
  $\langle\log R\rangle$   &     3.378 &     3.583 &     3.672 &     3.763 &     3.885 \\
  rms                      &     0.109 &     0.075 &     0.081 &     0.072 &     0.078 \\
% ***********************************************
\tableline\\[-9pt]
\multicolumn{6}{c}{{\em Group 3}  ~~ $n=2$ ~~ $\langle M_{R}\rangle_{\eta=2}=-22.26$}\\[2pt]
\tableline
% ***********************************************
  $\langle$SB$\rangle_{R}$ &     20.05 &     20.69 &     21.00 &     21.32 &     21.77 \\
  error                    & $\pm$0.31 & $\pm$0.35 & $\pm$0.37 & $\pm$0.43 & $\pm$0.49 \\ 
  $\langle\log R\rangle$   &     3.249 &     3.412 &     3.496 &     3.379 &     3.667 \\
  rms                      &     0.042 &     0.044 &     0.027 &     0.016 &     0.000 \\
% ***********************************************
\tableline\\[-9pt]
\multicolumn{6}{c}{{\em Grand mean}  ~~ $n=12/14$ ~~ $\langle M_{R}\rangle_{\eta=2}=-22.65$}\\[2pt]
\tableline
% ***********************************************
  $\langle$SB$\rangle_{R}$ &     20.60 &     21.31 &     21.69 &     22.08 &     22.50 \\
  error                    & $\pm$0.15 & $\pm$0.17 & $\pm$0.18 & $\pm$0.19 & $\pm$0.18 \\
  $\langle\log R\rangle$   &     3.394 &     3.594 &     3.692 &     3.789 &     3.890 \\
  rms                      &     0.146 &     0.151 &     0.163 &     0.168 &     0.160 \\
% ***********************************************
\enddata
% ***********************************************
\end{deluxetable}
% ***********************************************

% ******************************************************************
%  Table 14: Same as Tables 10 & 12 but for Cl 1604+4321.
% ******************************************************************
\begin{deluxetable}{ccccc}
\tablewidth{0pt}
\tabletypesize{\footnotesize}
\tablecaption{Same as Tables~\ref{tab:10} \& \ref{tab:12} 
              but for Cl 1604+4321.\label{tab:14}} 
% ***********************************************
\tablehead{
% ***********************************************
 \colhead{$\eta/\langle\log R\rangle$} &
 \colhead{4.104}                       & 
 \colhead{3.885}                       &
 \colhead{3.667}                       &
 \colhead{3.904}
\\
 \colhead{(1)}                         &  
 \colhead{(2)}                         &
 \colhead{(3)}                         & 
 \colhead{(4)}                         & 
 \colhead{(5)}
}     
% ***********************************************
\startdata
% ***********************************************
  1.0 &   19.16 & 18.64 & 18.27 & 18.74 \\
  1.3 &   19.76 & 19.24 & 18.87 & 19.34 \\
  1.5 &   20.13 & 19.61 & 19.24 & 19.71 \\
  1.7 &   20.44 & 19.92 & 19.57 & 20.04 \\
  2.0 &   20.84 & 20.32 & 19.97 & 20.44 \\
      &    G1   &  G2   &  G3   &  mean \\  
% ***********************************************
\enddata
% ***********************************************
\end{deluxetable}
% ***********************************************

% ******************************************************************
%  Table 15: Difference in the Surface Brightness Between the HST
%            Clusters and the Zero Redshift Standard SB vs. eta Curves
%            for Each of the Radius Bins at Each of the Fiducial eta
%            Values. 
% ******************************************************************
\begin{deluxetable}{ccccccccccc}
\tablewidth{0pt}
\tabletypesize{\footnotesize}
\tablecaption{Difference in the Surface Brightness Between the \textsl{HST} 
   Clusters and the Zero Redshift Standard SB vs.\ $\eta$ Curves for 
   Each of the Radius Bins at Each of the Fiducial $\eta$ 
   Values.\tablenotemark{a}\label{tab:15}} 
% ***********************************************
\tablehead{
% ***********************************************
 & \multicolumn{10}{c}{$\Delta\langle$SB$\rangle_{I}$ (mag)}
\\
                                      &
\multicolumn{3}{c}{Cl 1324+3011}      & &
\multicolumn{2}{c}{Cl 1604+4304}      & &
\multicolumn{3}{c}{Cl 1604+4321}       
\\
                                      &
\multicolumn{3}{c}{in $I_{\rm cape}$} & &
\multicolumn{2}{c}{in $I_{\rm cape}$} & &
\multicolumn{3}{c}{in $R_{\rm cape}$} 
\\
\cline{2-4}
\cline{6-7}
\cline{9-11}
 \colhead{$\eta$}                     &  
 \colhead{G1}                         &
 \colhead{G2}                         & 
 \colhead{G3}                         & &
 \colhead{G1}                         & 
 \colhead{G2}                         & &
 \colhead{G1}                         & 
 \colhead{G2}                         & 
 \colhead{G3}
\\
 \colhead{(1)}                        &  
 \colhead{(2)}                        &
 \colhead{(3)}                        & 
 \colhead{(4)}                        & &
 \colhead{(5)}                        & 
 \colhead{(6)}                        & &
 \colhead{(7)}                        & 
 \colhead{(8)}                        & 
 \colhead{(9)}
}     
% ***********************************************
\startdata
% ***********************************************
   1.0            &    1.69 & 1.87 & 1.96    &&       2.49 & 2.47        &&     1.87 & 1.85 & 1.78   \\
   1.3            &    2.20 & 1.93 & 2.13    &&       2.70 & 2.36        &&     2.03 & 1.97 & 1.82   \\
   1.5            &    2.43 & 1.94 & 2.04    &&       2.79 & 2.26        &&     2.16 & 1.94 & 1.76   \\
   1.7            &    2.64 & 1.96 & 1.88    &&       2.76 & 2.24        &&     2.34 & 2.00 & 1.75   \\
   2.0            &    2.68 & 1.99 & 1.88    &&       2.76 & 2.24        &&     2.38 & 2.08 & 1.80   \\
    n             &      2  &   6  &   5     &&         3  &   3         &&       3  &   7  &   2    \\
wt mean (mag)     & \multicolumn{3}{c}{2.04} && \multicolumn{2}{c}{2.51} && \multicolumn{3}{c}{1.99} \\
$(1+z)^{4}$ (mag) & \multicolumn{3}{c}{2.45} && \multicolumn{2}{c}{2.78} && \multicolumn{3}{c}{2.84} \\
% ***********************************************
\enddata
% ***********************************************
\tablenotetext{a}{The listings are the Tolman signal (in mags), degraded by  
   luminosity evolution in the look-back time.}
% ***********************************************
\end{deluxetable}
% ***********************************************

% ******************************************************************
%  Table 16: Average log r(eta)/r(2) vs. eta for the HST Clusters + Average 
% ******************************************************************
\begin{deluxetable}{crrrrrr}
\tablewidth{0pt}
\tabletypesize{\footnotesize}
\tablecaption{Average $\log r(\eta)/r(2)$ vs.\ $\eta$ Listed Separately 
             for Each of the \textsl{HST} Clusters and the Average Over
             All \textsl{HST} Galaxies. The Mean Values Are Shown in
             Figure~\ref{fig:15}.\label{tab:16}} 
% ***********************************************
\tablehead{
% ***********************************************
\multicolumn{7}{c}{$\langle\log r(\eta)/r(2)\rangle$}
\\
 \colhead{Cluster/$\eta$} &  
 \colhead{1.0}            &
 \colhead{1.3}            & 
 \colhead{1.5}            & 
 \colhead{1.7}            & 
 \colhead{2.0}            & 
 \colhead{n}
\\
 \colhead{(1)}            &  
 \colhead{(2)}            &
 \colhead{(3)}            & 
 \colhead{(4)}            & 
 \colhead{(5)}            & 
 \colhead{(6)}            & 
 \colhead{(7)}
}     
% ***********************************************
\startdata
% ***********************************************
   1324+3011   &   $-$0.489 & $-$0.291 & $-$0.195 & $-$0.091 & 0.000 &  11 \\
   1604+4304   &   $-$0.454 & $-$0.273 & $-$0.169 & $-$0.096 & 0.000 &   6 \\
   1604+4321   &   $-$0.510 & $-$0.310 & $-$0.213 & $-$0.115 & 0.000 &  13 \\[4pt]
      mean     &   $-$0.491 & $-$0.296 & $-$0.198 & $-$0.102 & 0.000 &  30 \\
% ***********************************************
\enddata
% ***********************************************
\end{deluxetable}
% ***********************************************

% ******************************************************************
% ***********              Figures                       ***********
% ******************************************************************
\clearpage

% ******************************************************************
%  Figure 1: Variation of Petrosian eta Radii with log r/r_e
% ******************************************************************
% ******************************************************************

% ******************************************************************
%  Figure 2: Relation Between log r/r_e and eta
% ******************************************************************
% ******************************************************************

% ******************************************************************
%  Figure 3: Correlation of eta with log r(eta)/r(2)
% ******************************************************************
% ******************************************************************

% ******************************************************************
%  Figure 4: Oemler profiles compared with observed T=-5 profiles
% ******************************************************************
% ******************************************************************

% ******************************************************************
%  Figure 5: Correlation of <SB> with Sersic n values
% ******************************************************************
% ******************************************************************

% ******************************************************************
%  Figure 6: Variation of the log r(eta=1)/r(2) radius ratio with 
%       intrinsic diameter in pc for the Virgo, Fornax, and Coma cluster
% ******************************************************************
% ******************************************************************

% ******************************************************************
%  Figure 7: Variation of the SB differences with intrinsic diameter
%            in pc for the Virgo, Fornax, and Coma cluster
% ******************************************************************
% ******************************************************************

% ******************************************************************
%  Figure 8: SB vs. eta for Cl 1324+3011
% ******************************************************************
% ******************************************************************

% ******************************************************************
%  Figure 9: data from Figure 8 plotted separately (Cl 1324+3011)
% ******************************************************************
% ******************************************************************

% ******************************************************************
%  Figure 10: SB vs. eta for Cl 1604+4304
% ******************************************************************
% ******************************************************************

% ******************************************************************
%  Figure 11: data from Figure 10 plotted separately (Cl 1604+4304)
% ******************************************************************
% ******************************************************************

% ******************************************************************
%  Figure 12: SB vs. eta for Cl 1604+4321
% ******************************************************************
% ******************************************************************

% ******************************************************************
%  Figure 13: data from Figure 12 plotted separately (Cl 1604+4321)
% ******************************************************************
% ******************************************************************

% ******************************************************************
%  Figure 14: log R(1)/R(2) vs. log R(2) for the HST cluster galaxies
% ******************************************************************
% ******************************************************************

% ******************************************************************
%  Figure 15: log r(eta)/r(2) vs. eta for the mean HST clusters
% ******************************************************************
% ******************************************************************

% ******************************************************************

\begin{thebibliography}{}
%
\bibitem[Binggeli \& Jerjen(1998)]{Binggeli:Jerjen:98}
   Binggeli, B., \& Jerjen H. 1998, 
   % Is the shape of the luminosity profile of dwarf elliptical
   % galaxies an useful distance indicator? 
   A\&A, 333, 17  
%
\bibitem[Binggeli, Sandage, \& Tarenghi(1984)Binggeli et~al.]{BST:84}
   Binggeli, B., Sandage, A., \& Tarenghi, M. 1984,
   % Studies of the Virgo Cluster. I - Photometry of 109 galaxies near
   % the cluster center to serve as standards 
   AJ, 89, 64
%
\bibitem[Bruzual \& Charlot(1993)]{Bruzual:Charlot:93}
   Bruzual, A.~G., \& Charlot, S. 1993, 
   % Spectral evolution of stellar populations using isochrone
   % synthesis 
   ApJ, 405, 538
%
\bibitem[BBC(1964)Burbidge, Burbidge, \& Crampin]{BBC:64}
   Burbidge, E.~M., Burbidge, G.~R., \& Crampin, D.~J. 1964, 
   % The Light Distribution in the Pair of Elliptical Galaxies 
   % NGC 4782-83  
   ApJ, 140, 1462
%
\bibitem[Caldwell(1987)]{Caldwell:87}
   Caldwell, N. 1987,
   % Dwarf elliptical galaxies in the Fornax cluster. I - A catalog
   % and luminosity function 
   AJ, 94, 1116
%
\bibitem[Caldwell \& Bothun(1987)]{Caldwell:Bothun:87}
   Caldwell, N., \& Bothun, G. 1987, 
   % Dwarf elliptical galaxies in the Fornax cluster. II - Their
   % structure and stellar populations 
   AJ, 94, 1126
%
\bibitem[Caon, Capaccioli, \& D'Onofrio(1993)Caon et~al.]{Caon:etal:93}
   Caon, N., Capaccioli, M., \& D'Onofrio, M. 1993,
   % On the Shape of the Light Profiles of Early Type Galaxies
   MNRAS, 265, 1013
%
\bibitem[Choloniewski(1985)]{Choloniewski:85}
   Choloniewski, J. 1985,
   % Bivariate luminosity function of E and S0 galaxies
   MNRAS, 214, 197
%
\bibitem[Code \& Welch(1979)]{Code:Welch:79}
   Code, A., \& Welch, G.~A. 1979,
   % Ultraviolet photometry from the Orbiting Astronomical
   % Observatory. XXVI - Energy distributions of seven early-type
   % galaxies and the central bulge of M31 
   ApJ, 228, 95 
%
\bibitem[Coleman et~al.(1980)Coleman, Wu, \& Weedman]{Coleman:etal:80}
   Coleman, G.~D., Wu, C.-C., \& Weedman, D.~W. 1980,
   % Colors and magnitudes predicted for high redshift galaxies
   ApJS, 43, 393
%
\bibitem[de~Vaucouleurs \& Longo(1988)]{deVaucouleurs:Longo:88}
   de~Vaucouleurs, A., \& Longo, G. 1988,
   Catalogue of Visual and Infrared Photometry of Galaxies from 0.5 $\mu$m
   to 10 $\mu$m 
   (Austin: Univ. of Texas) 
%
\bibitem[RC3(1991)de~Vaucouleurs et~al.]{RC3}
   de~Vaucouleurs, G., de~Vaucouleurs, A., Corwin, H.~G., Buta, R.~J., 
     Paturel, G., \& Fouqu{\'e}, P. 1991,
   Third Reference Catalogue of Bright Galaxies
   (New York: Springer) (RC3)
%
\bibitem[Djorgovski \& Davis(1987)]{Djorgovski:Davis:87}
   Djorgovski, S., \& Davis, M. 1987,
   % Fundamental properties of elliptical galaxies
   ApJ, 313, 59
%
\bibitem[Djorgovski \& Spinrad(1981)]{Djorgovski:Spinrad:81}
   Djorgovski, S.~B., \& Spinrad, H. 1981, 
   % Toward the application of metric size function in galactic
   % evolution and cosmology 
   ApJ, 251, 417
%
\bibitem[Ferguson \& Sandage(1988)]{Ferguson:Sandage:88}
   Ferguson, H.~C., \& Sandage, A. 1988, 
   % Population studies in groups and clusters of galaxies. I - The
   % luminosity function of galaxies in the Fornax Cluster 
   AJ, 96, 1520
%
\bibitem[Fraser(1977)]{Fraser:77}
   Fraser, C.~W. 1977,
   % Photographic surface photometry of galaxies in the Virgo Cluster
   A\&AS, 29, 161
%
\bibitem[Greenstein(1938)]{Greenstein:38}
   Greenstein, J.~L. 1938, 
   % The Temperatures of the Extragalactic Nebulae and the Redshift
   % Correction 
   ApJ, 88, 605
%
\bibitem[Hodge(1978)]{Hodge:78}
   Hodge, P.~M. 1978, 
   % Surface photometry of bright members of the Fornax Cluster of
   % galaxies 
   ApJS, 37, 429 
%
\bibitem[Hubble(1934)]{Hubble:34}
   Hubble, E. 1934, 
   % The Distribution of Extra-Galactic Nebulae
   ApJ, 79, 8 
%
\bibitem[Hubble(1936)]{Hubble:36}
   Hubble, E. 1936, 
   % Effects of Red Shifts on the Distribution of Nebulae
   ApJ, 84, 517
%
\bibitem[Hubble(1937)]{Hubble:37}
   Hubble, E. 1937, 
   The Observational Approach to Cosmology
   (Oxford: Clarendon Press)
%
\bibitem[Hubble(1953)]{Hubble:53}
   Hubble E. 1953, 
   % The law of red shifts (George Darwin Lecture)
   MNRAS, 113, 658, {\em The Darwin Lecture}
%
\bibitem[Hubble \& Tolman(1935)]{Hubble:Tolman:35}
   Hubble, E., \& Tolman, R. C. 1935, 
   % Two Methods of Investigating the Nature of the Nebular Redshift
   ApJ, 82, 302 
%
\bibitem[Ichikawa, Wakamatsu, \& Okamura(1986)Ichikawa et~al.]{Ichikawa:etal:86}
   Ichikawa, S.-I., Wakamatsu, K.-I., \& Okamura, S. 1986,
   % Surface photometry of dwarf elliptical galaxies in the Virgo cluster
   ApJS, 60, 475
%
\bibitem[Impey, Bothun, \& Malin(1988)Impey et~al.]{Impey:etal:88}
   Impey, C., Bothun, G. \& Malin, D. 1988, 
   % Virgo dwarfs - New light on faint galaxies
   ApJ, 330, 634
%
\bibitem[Jedrzejewski(1987)]{Jedrzejewski:87}
   Jedrzejewski, R. 1987, 
   % CCD surface photometry of elliptical galaxies. I - Observations,
   % reduction and results 
   MNRAS, 226, 747
%
\bibitem[Johnson(1965)]{Johnson:65}
   Johnson, H.~L. 1965, 
   % Infrared Photometry of M-Dwarf Stars
   ApJ, 141, 170
%
\bibitem[King(1978)]{King:78}
   King, I. 1978,
   % Surface photometry of elliptical galaxies
   ApJ, 222, 1
%
\bibitem[Kormendy(1977)]{Kormendy:77}
   Kormendy, J. 1977,
   % Brightness distributions in compact and normal galaxies. II -
   % Structure parameters of the spheroidal component 
   ApJ, 218, 333
%
\bibitem[Kormendy(1987)]{Kormendy:87}
   Kormendy, J. 1987, 
   % Cores of early-type galaxies - The nature of dwarf spheroidal galaxies
   in Nearly Normal Galaxies: From the Planck time to the present, 
   ed. S.~Faber (New York: Springer), 163
%
\bibitem[Kormendy et~al.(2009)]{Kormendy:etal:09}
   Kormendy, J., Fisher, D.~B., Cornell, M.~E., \& Bender, R. 2009,
   %  Structure and Formation of Elliptical and Spheroidal Galaxies
   ApJS, 182, 216  % arXiv:0810.1681
%
\bibitem[Kron(1995)]{Kron:95}
   Kron, R.~G. 1995,
   % Evolution in the Galaxy Population
   in Saas-Fee Advanced Course 23, The Deep Universe, 
   ed. B.~Binggeli \& R.~Buser 
   (Berlin: Springer), 233
%
\bibitem[Landolt(1983)]{Landolt:83}
   Landolt, A. 1983, 
   % UBVRI photometric standard stars around the celestial equator
   AJ 88, 489 (UBVRI)
%
\bibitem[Landolt(1992)]{Landolt:92}
   Landolt, A. 1992,
   % UBVRI photometric standard stars in the magnitude range 11.5-16.0
   % around the celestial equator 
   AJ, 104, 340 (UBVRI)
%
\bibitem[Lubin et~al.(1998)]{LPO:98}
   Lubin, L.~M., Postman, M., Oke, J.~B., Ratnatunga, K.~U., 
   Gunn, J.~E., Hoessel, J.~G., \& Schneider, D.~P. 1998,
   % A Study of Nine High-Redshift Clusters of Galaxies. III. Hubble
   % Space Telescope Morphology of Clusters 0023+0423 and 1604+4304 
   AJ, 116, 584 
%
\bibitem[Lubin et~al.(2001)]{LPO:01}
   Lubin, L.~M., Postman, M., Oke, J.~B., Brunner, R., Gunn, J.~E., \& 
   Schneider, D. 2001, 
   % 
   AJ, said to be in preparation 
%
\bibitem[LS01a(2001a)Sandage \& Lubin]{LS:01a}
   Lubin, L.~M., \& Sandage, A. 2001a, 
   % The Tolman Surface Brightness Test for the Reality of the
   % Expansion. II. The Effect of the Point-Spread Function and Galaxy
   % Ellipticity on the Derived Photometric Parameters 
   AJ, 121, 2289 (LS01a), Paper~II 
%
\bibitem[LS01b(2001b)Sandage \& Lubin]{LS:01b}
   Lubin, L.~M., \& Sandage, A. 2001b, 
   % The Tolman Surface Brightness Test for the Reality of the
   % Expansion. III. Hubble Space Telescope Profile and Surface
   % Brightness Data for Early-Type Galaxies in Three High-Redshift
   % Clusters 
   AJ, 122, 1071, (LS01b), Paper~III 
%
\bibitem[LS01c(2001c)Sandage \& Lubin]{LS:01c}
   Lubin, L.~M., \& Sandage, A. 2001c, 
   % The Tolman Surface Brightness Test for the Reality of the
   % Expansion. IV. A Measurement of the Tolman Signal and the
   % Luminosity Evolution of Early-Type Galaxies 
   AJ, 122, 1084, (LS01c), Paper~IV 
%
\bibitem[Malumuth \& Kirshner(1985)]{Malumuth:Kirshner:85}
   Malumuth, E.~M., \& Kirshner, R.~P. 1985,
   % Dynamics of luminous galaxies. II - Surface photometry and
   % velocity dispersions of brightest cluster members 
   ApJ, 291, 8
%
\bibitem[Mattig(1958)]{Mattig:58}
   Mattig, W. 1958, 
   % Über den Zusammenhang zwischen Rotverschiebung und scheinbarer
   % Helligkeit 
   Astron. Nachr., 284, 109
%
\bibitem[Michard(1979)]{Michard:79}
   Michard, R. 1979,
   % 'Sequences' and 'populations' of early-type galaxies
   A\&A, 74, 206
%
\bibitem[Molaro et~al.(2002)]{Molaro:etal:02}
   Molaro, P., Levshakov, S.~A., Dessauges-Zavadsky, M., \& D'Odorico, S. 2002, 
   % The cosmic microwave background radiation temperature at 
   % zabs=3.025 toward QSO 0347-3819 
   A\&A, 381, L64
%
\bibitem[Oberth(1923)]{Oberth:23}
   Oberth, H. 1923,
   Die Rakete zu den Planetenr{\"a}umen 
   (M{\"u}nchen: Oldenbourg Verlag)
%
\bibitem[Oemler(1973)]{Oemler:73}
   Oemler, A. 1973, 
   % The Systematic Properties of Clusters of Galaxies
   Ph.D. thesis, California Institute of Technology
%
\bibitem[Oemler(1974)]{Oemler:74}
   Oemler, A. 1974,
   % The Systematic Properties of Clusters of Galaxies. Photometry of
   % 15 Clusters 
   ApJ, 194, 1
%
\bibitem[Oemler(1976)]{Oemler:76}
   Oemler, A. 1976, 
   % The structure of elliptical and cD galaxies
   ApJ, 209, 693
%
\bibitem[OPL(1998)Oke, Postman, \& Lubin]{OPL:98}
   Oke, J.~B., Postman, M., \& Lubin, L.~M. 1998, 
   % A Study of Nine High-Redshift Clusters of Galaxies. I. The Survey
   AJ, 116, 549 (OPL\,1998)
%
\bibitem[Oke \& Sandage(1968)]{Oke:Sandage:68}
   Oke, J.~B., \& Sandage, A. 1968,
   % Energy Distributions, K Corrections, and the Stebbins-Whitford
   % Effect for Giant Elliptical Galaxies 
   ApJ, 154, 21
%
\bibitem[Pence(1976)]{Pence:76}
   Pence, W. 1976,
   % K-corrections for galaxies of different morphological types
   ApJ, 203, 39
%
\bibitem[Petrosian(1976)]{Petrosian:76}
   Petrosian, V. 1976, 
   % Surface brightness and evolution of galaxies
   ApJ, 209, L1
%
\bibitem[PL95(1995)Postman \& Lauer]{PL:95}
   Postman, M., \& Lauer, T. 1995, 
   % Brightest cluster galaxies as standard candles
   ApJ, 440, 28 (PL95) 
%
\bibitem[PLO(1998)Oke, Postman, \& Lubin]{PLO:98}
   Postman, M., Lubin, L., \& Oke, J.~B. 1998, 
   % A Study of Nine High-Redshift Clusters of
   % Galaxies. II. Photometry, Spectra, and Ages of Clusters 0023+0423
   % and 1604+4304 
   AJ, 116, 560 (PLO\,1998)
%
\bibitem[PLO(2001)Oke, Postman, \& Lubin]{PLO:01}
   Postman, M., Lubin, L., \& Oke, J.~B. 2001,
   % A Study of Nine High-Redshift Clusters of
   % Galaxies. IV. Photometry and Spectra of Clusters 1324+3011 and
   % 1604+4321 
   AJ, 122, 1125 (PLO\,2001)
%
\bibitem[Poulain \& Nieto(1994)]{Poulain:Nieto:94}
   Poulain, P., \& Nieto, J-L. 1994, 
   % UBVRI photoelectric photometry of bright southern early-type galaxies
   A\&AS, 103, 573
%
\bibitem[Sandage(1961)]{Sandage:61}
   Sandage, A. 1961,
   % The Ability of the 200-INCH Telescope to Discriminate Between
   % Selected World Models. 
   ApJ, 133, 355
%
\bibitem[Sandage(1968)]{Sandage:68}
   Sandage, A. 1968, 
   % The Time Scale for Creation
   in Galaxies and the Universe, 
   ed. L. Woltjer 
   (New York: Columbia Univ. Press), 75
%
\bibitem[Sandage(1972)]{Sandage:72}
   Sandage, A. 1972, 
   % The Redshift-Distance Relation. I. Angular Diameter of First
   % Ranked Cluster Galaxies as a Function of Redshift: the Aperture
   % Correction to Magnitudes 
   ApJ, 173, 485 
%
\bibitem[Sandage(1973)]{Sandage:73}
   Sandage, A. 1973,
   % The Redshift-Distance Relation.VI. The Hubble Diagram from S20
   % Photometry for Rich Clusters and Sparse Groups: a Study of
   % Residuals 
   ApJ, 183, 731
%
\bibitem[Sandage(1974)]{Sandage:74}
   Sandage, A. 1974, 
   % Cosmology with the LST
   in Large Space Telescope -- A New Tool For Science,
   ed. P.~F. Simmons
   (New York: AIAA), 19
%
\bibitem[Sandage(1988)]{Sandage:88}
   Sandage, A. 1988,
   % Observational tests of world models
   ARA\&A, 26, 561
%
\bibitem[Sandage(1995)]{Sandage:95}
   Sandage, A. 1995, 
   % Practical Cosmology: Inventing the Past
   in Saas-Fee Advanced Course 23, The Deep Universe, 
   ed. B.~Binggeli \& R.~Buser 
   (Berlin: Springer), 1 % 1--232
%
\bibitem[Sandage(1997)]{Sandage:97}
   Sandage, A. 1997,
   % The Mount Wilson Halo Mapping Project 1975-1985 I: The
   % UBV(RI)_M_W Photometric System Compared with Other Standard
   % Systems: The Adopted Trigonometric HR Diagram in (R-I)_M_W and
   % (V-I)_M_W 
   PASP, 109, 1193
%
\bibitem[Sandage(1998)]{Sandage:98}
   Sandage, A. 1998, 
   % Beginnings of Observational Cosmology in Hubble's Time:
   % Historical Overview
   in STScI Symp. 11, The Hubble Deep Field, 
   ed. M.~Livio, S.~M. Fall, \& P.~Madau
   (New York: Cambridge Univ. Press), 1 
%
\bibitem[Sandage(2001)]{Sandage:01}
   Sandage, A. 2001,
   % The Mount Wilson Halo Mapping Project 1975-1985. II. Photometric
   % Properties of the Mount Wilson Catalogue of Photographic Magnitudes
   % in Selected Areas 1-139 
   PASP, 113, 267
%
\bibitem[SL01(2001)Sandage \& Lubin]{SL:01}
   Sandage, A., \& Lubin, L.~M. 2001,
   % The Tolman Surface Brightness Test for the Reality of the Expansion. 
   % I. Calibration of the Necessary Local Parameters
   AJ, 121, 2271 (SL01), Paper~I
%
\bibitem[SP90a(1990a)Sandage \& Perelmuter]{SP:90a}
   Sandage, A., \& Perelmuter, J.-M. 1990a, 
   % The surface brightness test for the expansion of the universe. I
   % - Properties of Petrosian metric diameters 
   ApJ, 350, 481 (SP90a)
%
\bibitem[SP90b(1990b)Sandage \& Perelmuter]{SP:90b}
   Sandage, A., \& Perelmuter, J.-M. 1990b, 
   % The surface brightness test for the expansion of the universe. II
   % - Radii, surface brightness, and absolute magnitude correlations
   % for nearby E galaxies 
   ApJ, 361, 1 (SP90b)
%
\bibitem[SP91(1991)Sandage \& Perelmuter]{SP:91}
   Sandage, A., \& Perelmuter, J.-M. 1991, 
   % The surface brightness test for the expansion of the
   % universe. III - Reduction of data for the several brightest
   % galaxies in clusters to standard conditions and a first indication
   % that the expansion is real 
   ApJ, 370, 455 (SP91)
%
\bibitem[Sandage \& Smith(1963)]{Sandage:Smith:63}
   Sandage, A., \& Smith, L.~L. 1963, 
   % A four-color photometric system applied to line blanketing of
   % subdwarfs 
   ApJ, 137, 1057
%
\bibitem[Schild \& Oke(1971)]{Schild:Oke:71}
   Schild, R., \& Oke, J.~B. 1971, 
   % Energy Distributions and K-Corrections for the Total Light from
   % Giant Elliptical Galaxies 
   ApJ, 169, 209
%
\bibitem[Schombert(1986)]{Schombert:86}
   Schombert, J.~M. 1986, 
   % The structure of brightest cluster members. I - Surface photometry
   ApJS, 60, 603
%
\bibitem[Schombert(1987)]{Schombert:87}
   Schombert, J.~M. 1987,
   % The structure of brightest cluster members. II - Mergers
   ApJS, 64, 643
%
\bibitem[Seares, Kapteyn, \& van Rhijn(1930)Seares et~al.]{Seares:etal:30}
   Seares, F.~H., Kapteyn, J.~C., \& van~Rhijn, P.~J. 1930,
   The Mount Wilson Catalogue of Photographic Magnitudes in Selected
   Areas 1-39, Publ. 402
   (Washington: Carnegie Inst. of Washington)
%
\bibitem[S{\'e}rsic(1968)]{Sersic:68}
   S{\'e}rsic, J-L. 1968,
   Atlas de Galaxias Australes
   (C{\'o}rdoba: Obs. Astron.)
%
\bibitem[Songaila et~al.(1994)]{Songaila:etal:94}
   % Songaila, A., Cowie, L.~L., Vogt, S., Keane, M., Wolfe, A.~M.,
   % Hu, E.~M., Oren, A.~L., Tytler, D.~R., Lanzetta, K.~M. 
   Songaila, A., et~al. 1994, 
   % Measurement of the Microwave Background Temperature at a Redshift
   % % of 1.776 
   Nature, 371, 43 
%
\bibitem[Spitzer(1946)]{Spitzer:46}
   Spitzer, L. 1946, 
   % The Astronomical Advantages of an Extraterrestrial Observatory
   RAND project report, reprinted in Astron. Quart., 7, 131 (1990) 
%
\bibitem[Srianand, Petitjean, \& Ledoux(2000)Srianand et~al.]{Srianand:etal:00}
   Srianand, R., Petitjean, P., \& Ledoux, C. 2000, 
   % The cosmic microwave background radiation temperature at a
   % redshift of 2.34 
   Nature, 408, 931
%
\bibitem[Stebbins, Whitford, \& Johnson(1950)Stebbins et~al.]{Stebbins:etal:50}
   Stebbins, J., Whitford, A.~E., \& Johnson, H.~L. 1950,
   % Photoelectric Magnitudes and Colors of Stars in Selected Areas
   % 57, 61, and 68. 
   ApJ, 112, 469
%
\bibitem[Strom \& Strom(1978a)]{Strom:Strom:78a}
   Strom, K.~M., \& Strom, S.~E. 1978a,
   % Surface brightness and color distributions of elliptical and S0
   % galaxies. I - The Coma cluster elliptical galaxies 
   AJ, 83, 73
%
\bibitem[Strom \& Strom(1978c)]{Strom:Strom:78c}
   Strom, K.~M., \& Strom, S.~E. 1978c,
   % Surface brightness and color distributions of elliptical and S0
   % galaxies. III - E galaxies in the clusters Abell 1228, Abell 2151
   % /Hercules/, and Abell 2199 
   AJ, 83, 1293 
%
\bibitem[Strom \& Strom(1978b)]{Strom:Strom:78b}
   Strom, S.~E., \& Strom, K.~M. 1978b, 
   % Surface brightness and color distributions of elliptical and SO
   % galaxies. II - E galaxies in the clusters Abell 426 /Perseus/ and
   % Abell 1367 
   AJ, 83, 732
%
\bibitem[Thomsen \& Frandsen(1983)]{Thomsen:Frandsen:83}
   Thomsen, B. \& Frandsen, S. 1983,
   % Surface brightness and effective radius for elliptical galaxies
   AJ, 88, 789
%
\bibitem[Thuan \& Romanishin(1981)]{Thuan:Romanishin:81}
   Thuan, T.~X., \& Romanishin, W. 1981,
   % The structure of giant elliptical galaxies in poor clusters of
   % galaxies 
   ApJ, 248, 439
%
\bibitem[Tolman(1930)]{Tolman:30}
   Tolman, R.~C. 1930, 
   % On the Estimation of Distances in a Curved Universe with a
   % Non-Static Line Element 
   Proc. Nat. Acad. Sci., 16, 511
%
\bibitem[Tolman(1934)]{Tolman:34}
   Tolman, R.~C. 1934,
   Relativity, Thermodynamics, \& Cosmology 
   (Oxford: Clarendon Press), 467  
%
\bibitem[Vigroux et~al.(1988)]{Vigroux:etal:88}
   Vigroux, L., Souviron, J., Lachieze-Rey, M., \& Vader, J.~P. 1988, 
   % Three-color surface photometry of a selected sample of early-type
   % galaxies. I - Observations and data reduction 
   A\&AS, 73, 1
%
\bibitem[Wells(1973)]{Wells:73}
   Wells, D.~C. 1973,
   % Integrated Spectral Energy Distributions of Galaxies
   Ph.D. thesis, Univ. of Texas at Austin %  Univ. Texas Pub. 13
%
\bibitem[Whitford(1971)]{Whitford:71}
   Whitford, A.~E. 1971, 
   % Absolute Energy Curves and K-Corrections for Giant Elliptical Galaxies
   ApJ, 169, 215
%
\bibitem[Yoshii \& Takahara(1988)]{Yoshii:Takahara:88}
   Yoshii, Y., \& Takahara, F. 1988,
   % Galactic evolution and cosmology - Probing the cosmological
   % deceleration parameter 
   ApJ, 326, 1
%
% ******************************************************************
\end{thebibliography}
\end{document}